\documentclass[useAMS,usenatbib,a4paper]{mn2e}
\newcommand {\aplt} {{\raise-.5ex\hbox{$\buildrel<\over\sim$}}} 
\usepackage{amssymb,amsmath,graphicx}
\usepackage{url,lscape}

 \title[TESS stellar parameters]{Data-driven stellar parameters for southern TESS FGK targets}

   \author[N.R.\ Deacon et al.]{N.R. Deacon$^{1}$, Th. Henning$^{1}$, D.E. Kossakowski$^{1}$  \\
   $^1$Max Planck Institute for Astronomy, Konigstuhl 17, Heidelberg, 69117, Germany\\
             }
 
\begin{document}
 \date{}
 \pagerange{\pageref{firstpage}--\pageref{lastpage}} \pubyear{2015}
 \maketitle
 \label{firstpage}
  \begin{abstract}
We present stellar parameter estimates for  939,457 southern FGK stars that are candidate targets for the TESS mission. Using a data-driven method similar to the CANNON, we build a model of stellar colours as a function of stellar parameters. We then use these in combination with stellar evolution models to estimate the effective temperature, gravity, metallicity, mass, radius and extinction for our selected targets. Our effective temperature estimates compare well with those from spectroscopic surveys and the addition of Gaia DR2 parallaxes allows us to identify subgiant interlopers into the TESS sample. We are able to estimate the radii of TESS targets with a typical uncertainty of 9.3\%. This catalogue can be used to screen exoplanet candidates from TESS and provides a homogeneous set of stellar parameters for statistical studies.
  \end{abstract}
%

\section{Introduction}
Dedicated space missions studying planets around other stars have been one of the scientific stories of the decade. The recently concluded NASA Kepler mission \citep{Borucki2010} was a huge leap in the study of exoplanets providing both fascinating new worlds to characterise and unprecedented statistical power, unveiling over 3,000 planet candidates for further studies. Kepler's successor is the Transiting Exoplanet Survey Satellite (TESS; \citealt{Ricker2015}). This is currently surveying the sky as part of its two year mission. For the first year it will survey the ecliptic southern hemisphere. Full frame images (i.e. long-cadence photometry) will be taken every 30 minutes, whereas a select 200,000 targets will have short-cadence photometry every 2 minutes.

Stellar parameters are a key ingredient in the study of transiting exoplanets. The most crucial of these is stellar radius. This is because the depth of the dip in a transit lightcurve depends on the ratio between the planet's radius and the star's radius. Hence the measurement of a transiting planet's radius requires a reliable estimate of its host star's radius. This dependence on stellar radius incentivises exoplanet surveys to observe dwarf stars, especially cool dwarfs as they will have smaller planets for any given transit depth. Hence giants and subgiants are often treated as contaminants in target lists for such surveys.

The Kepler Input Catalog \citep{Brown2011} provided stellar parameter estimates for all Kepler targets. These were based largely on photometry with dwarf-giant separation being done using colours for most stars. The recent availability of parallaxes from Data Release 2 (DR2) of the ESA Gaia mission \citep{Prusti2016} has led to a retrospective redetermination of the stellar parameters for Kepler targets (\citealt{Berger2018} based on the methods developed by \citealt{Huber2017}). This therefore produced significant changes in the estimation of the radius distribution of exoplanets discovered by Kepler. 

The main stellar characterisation preparatory work for TESS was carried out by \cite{Stassun2018}. This detailed the production of two catalogues, the TESS Input Catalog (TIC) and the Candidate Target List (CTL). In brief, the TESS Input Catalog is a compendium of information about every single source that may fall on a TESS pixel. This helps provide information on flux contamination for TESS target but also feeds into the smaller CTL. The CTL is a list of almost four million sources which are each a candidate to be one of the $\sim$ 200,000 TESS short-cadence targets. It contains stellar parameter estimates along with a priority statistic which is intended to estimate an object's suitability to be a TESS short cadence target. While the exact priority algorithm is not explicitly published in \cite{Stassun2018}, it contains an extensive description of the factors that go into its calculation and a formula for the priorities' calculation referencing a work in preparation for full calculation. 

The stellar parameters in the TESS CTL are based on the astronomical surveys that were available when it was compiled. For stars covered by large-scale spectroscopic surveys such as APOGEE \citep{Zasowski2017} and LAMOST \citep{Luo2015} spectroscopically determined stellar parameters were used. For the majority of stars lacking these observations, colours were used to determine effective temperature and reddening. For stars with a parallax, the radius was calculated using a bolometric correction and absolute magnitudes derived using this parallax. For stars lacking a parallax a temperature-radius relationship derived from eclipsing binaries was used. This latter relationship relies on the objects in question being dwarfs. As the CTL was constructed prior to Gaia DR2, most stars listed do not have measured parallaxes. Due to this, reduced proper motion cuts were used to exclude giants from the CTL sample. The CTL prioritises bright stars (for ease of observation) and small, cool stars (to maximise the transit depth of small planets). This leads to a bimodal distribution in temperature with the vast majority of targets falling the ranges 4800\,K$<T_{eff}<$7000\,K and 3000\,K$<T_{eff}<$4000\,K.

In this paper we aim to provide stellar parameters for southern TESS FGK targets. We choose to limit ourselves to this subsample of TESS targets for two reasons. Firstly, we wish to use the metallicity-sensitive blue bands from SkyMapper \citep{Wolf2018} which only cover the southern hemisphere. Secondly, we lack a good quality training sample for southern M dwarfs. Our method includes Gaia~DR2 data and thus we have many more stellar distance measurements available to us than \cite{Stassun2018} did when producing their catalogue. These distances will allow us to better determine the radii of TESS targets.

\section{Methods}
\subsection{Observational data}
Our aim for this project is to measure the stellar parameters for FGK stars that are possible TESS targets in the CTL. These are typically bright objects ($G \lesssim 14$\,mag). Hence many TESS targets will be saturated in optical surveys such as Pan-STARRS\,1 \citep{Chambers2016} or SDSS \citep{SDSS_DR9}. While infrared surveys such as 2MASS \citep{Skrutskie2006} or WISE \citep{Cutri2013} will have significantly fewer saturated TESS targets, these surveys do not have the same sensitivity to metallicity that blue-optical surveys do. Hence we choose to make use of the SkyMapper survey \citep{Wolf2018}. This covers the southern hemisphere, albeit with some gaps in the coverage of the Galactic Plane. SkyMapper is significantly shallower than Pan-STARRS\,1 or SDSS and thus suffers less from saturation. It also has two blue-optical bands $u_{skymapper}$ and $v_{skymapper}$ which are sensitive to metallicity. We choose to exclude the $g_{skymapper}$ and $r_{skymapper}$ bands from our analysis as these are deeper and have fainter saturation levels than the other SkyMapper bands. We use all three ($J$, $H$, $K_S$) bands from 2MASS and the $W1$ and $W2$ bands from WISE. The two redder WISE bands ($W3$, $W4$) are shallow, have lower spatial resolution and would only add a small amount of information about the tail of the SEDs for our target stars. Hence we have nine photometric datapoints ($u_{skymapper}$, $v_{skymapper}$, $i_{skymapper}$, $z_{skymapper}$, $J$, $H$, $K_S$, $W1$ and $W2$) across our SEDs. We also make use of Gaia DR2 data \citep{Prusti2016,Brown2018} using both the Gaia $G$ band magnitude and parallax. 

For our training sample we assume that the objects are not moving and thus do not correct for proper motion when searching for counterparts in Gaia, 2MASS, WISE or Skymapper. For objects in the TESS CTL we use the Gaia matching process outlined in Appendix~\ref{gaia_section}. We then use the Gaia epoch=2015.5 position when searching for Skymapper. We use the CTL values for 2MASS and an epoch of  2010.5 for WISE. We set our search radius as one arcsecond or 1.5 times the total proper motion, whichever is greater. This allows us to account for the movement of the highest proper motion star. Appendix~\ref{gaia_section} contains a table of matched Gaia objects for 3,817,768 of the 3,848,012 objects in the TESS CTL. Any object not listed in this table does not have a Gaia match from our algorithm. Appendix~\ref{gaia_section} also contains a detailed description of our GAIA matching algorithm.

For the distance values for each star in our TESS target list we used an inversion of the measured {\it Gaia} parallax and a simple Taylor expansion to propagate the parallax errors to the absolute magnitude estimate. For stars with low-significance parallaxes ($\frac{\sigma_{\pi}}{\pi}>0.1$) this would not produce accurate estimates of the absolute magnitude or its uncertainty \citep{Bailer-Jones2015}. However as the vast majority of our TESS target stars will be both bright and close, the parallaxes for the vast majority of our targets will be of a high enough significance for this not to be a problem. We exclude objects with noisy astrometric solutions by requiring the Reduced Unit-Weighted Error is below 1.5 \citep{Lindegren2018,Lindegren2018a}.  We also use Equation 2 from \cite{Lindegren2018a} to calculate our parallax uncertainties. This is a scaling of the internal parallax errors by 1.08 added in quadrature to an additional noise floor of 0.021\,milliarcseconds for bright stars ($G<$13) or 0.043\,milliarcseconds for faint stars ($G>13$).

We applied a number of data quality cuts to ensure we only used accurate photometry. Firstly, we ignore SkyMapper magnitudes where an object was not well detected in each observation in that filter (requiring NGOOD=NVISIT), where there were no good individual visit detections or where that filter had data warning flags set (requiring that FLAGS and NIMAFLAGS were zero). We also excluded 2MASS and WISE magnitudes where the objects were not flagged as having "A" photometric quality in a particular band.
 
\subsection{Statistical method}
 Our method combines evolutionary models with data-driven models of stellar colours. We employ a Markov Chain Monte Carlo method (MCMC) to generate our stellar parameters. For each star we explore a space defined by initial mass, age and metallicity. For each step in the chain, we use the {\it PARSEC} stellar models of \citep{Marigo2017} to generate effective temperatures, surface gravities and absolute magnitudes for the star and then use a data-driven method to generate colours. 

We used a variation on the {\it CANNON} statistical method laid out by \cite{Ness2015}. In this method each star has a series of measurements (i.e. observables) and labels (i.e. stellar parameters). The original {\it CANNON} method used normalised spectra and hence its measurements are the relative flux values at different wavelengths. For our measurements we use colours. All of our colours contain the GAIA $G$ band and so we are in effect applying the {\it CANNON} to an extremely low resolution, normalised spectrum. We use nine colours, four using {\it Skymapper} filters ($u_{skymapper}-G_{Gaia}$, $v_{skymapper}-G_{Gaia}$, $i_{skymapper}-G_{Gaia}$, $z_{skymapper}-G_{Gaia}$), three using 2MASS filters ($G_{Gaia}-J_{2MASS}$, $G_{Gaia}-H_{2MASS}$, $G_{Gaia}-K_{S,2MASS}$) and two derived from {\it WISE} photometry ($G_{Gaia}-W1$, $G_{Gaia}-W2$). As we will see later all of these colours are sensitive to temperature with the two bluest ($u_{skymapper}-G_{Gaia}$, $v_{skymapper}-G_{Gaia}$) being most sensitive to metallicity. The seven other colours also show sensitivity to gravity at temperatures below 5000\,K. 

Say we have $k$ stars each with $j$ labels and $i$ measurements. The likelihood will be,
\begin{equation}
ln L = \sum\left (-\frac{1}{2}\frac{(meas_{i,k}-f(a_{i,j},label_{j,k}))^2}{\sigma_{i,k}^2+s_i^2} -\frac{1}{2}ln(\sigma_{i,k}^2+s_i^2)\right )
\label{like1}
\end{equation} 
Where the parameters $a_{i,j}$ are our model coefficients, $s_{i}$ is the scatter term fitted by our model, $meas_{i,k}$ are the individual measurements (colours or magnitudes), $\sigma_{i,k}$ the uncertainties on those measurements and $label{j,k}$ are the stellar parameter labels. Our scatter term is in-effect all the possible stellar parameters (abundances, binarity, etc.) that are not included in our model packaged up into one term. We use three labels ($[Fe/H]$, $\log g$ and $\theta$) where $\theta=\frac{5040 \text{\,K}}{T_{eff}}$ to describe the factors that could affect the colours of each object. The function we choose to model each measurement (i.e. colour) is fourth order in $\theta$, first order the other two labels.
\begin{equation}
\begin{aligned}
model_{i,k}=\sum_j f(a_{i,j},label_{j,k})=&a_0+a_1\theta+a_2\theta^2+a_3\theta^3+a_4\theta^4\\
&+a_5\log g+a_6[Fe/H]
\end{aligned}
\label{model}
\end{equation}
To train our method we used a set of stars with stellar parameters determined from the {\it GALAH} survey \citep{Buder2018}. Using the known labels from this survey we calculated a series of $i$ likelihoods, one for each measurement (in our case colour) that we have. We then maximised these likelihoods over all the stars in our sample to determine the parameters set $a_{i,j}$
\begin{equation}
ln L = \sum_{j,k}\left (-\frac{1}{2}\frac{(meas_{i,k}-f(a_{i,j},label_{j,k}))^2}{\sigma_{i,k}^2+s_i^2} -\frac{1}{2}ln(\sigma_{i,k}^2+s_i^2)\right )
\label{like2}
\end{equation}
We then used  these parameters to find the best-fit values of each of the $j$ labels for each star $k$. 
\begin{equation}
ln L = \sum_{i,j}\left (-\frac{1}{2}\frac{(meas_{i,k}-f(a_{i,j},label_{j,k}))^2}{\sigma_{i,k}^2+s_i^2} -\frac{1}{2}ln(\sigma_{i,k}^2+s_i^2)\right )
\label{like3}
\end{equation}
We also have another measurement to compare to, the absolute $G$-band measurement from {\it Gaia}. The comparison value for this comes directly from the {\it PARSEC} stellar models and thus comes with no scatter. To include an estimate of the scatter on the absolute $G$ magnitude we extracted a sample of stars around the Praesepe cluster. We then selected cluster members having proper motions within 3\,mas/yr of a by-eye estimate of the mean cluster proper motion ($\mu_{\alpha}\cos\delta=-36$\,mas/yr, $\mu_{\delta}=-13$\,mas/yr) and with parallaxes between 4.5 and 6.0 milliarcseconds. We then compared this cluster population with a Praesepe-age (790\,Myr, \citealt{Brandt2015}) {\it PARSEC} isochrone. After excluding stars that appeared to have evolved (absolute $G$ magnitude $<$1) and late-type stars ($G_{Gaia}-K_{S,2MASS}>$3) we calculated the difference between the absolute $G$ magnitude calculated using the Gaia DR2 parallax and the expected absolute $G$ magnitude given the star's $G_{Gaia}-K_{S,2MASS}$ colour and the {\it PARSEC} isochrone. We used a simple linear interpolation between points in the isochrone in $\log_{10}mass$, $\log_{10}age$ and [Fe/H]. We then estimated the scatter in absolute $G$ magnitude by taking the median absolute deviation of these magnitude differences and multiplying by 1.48. The median absolute deviation is robust to outliers. The factor of 1.48 is the scale factor to transform the median absolute deviation to a 1$\sigma$ scatter estimate \citep{Maronna2006}. This gives us an estimate of the scatter of 0.153 magnitudes that is robust to outliers. 

\subsection{Training the model}
\begin{figure*}
 \setlength{\unitlength}{1mm}
\includegraphics[scale=0.55]{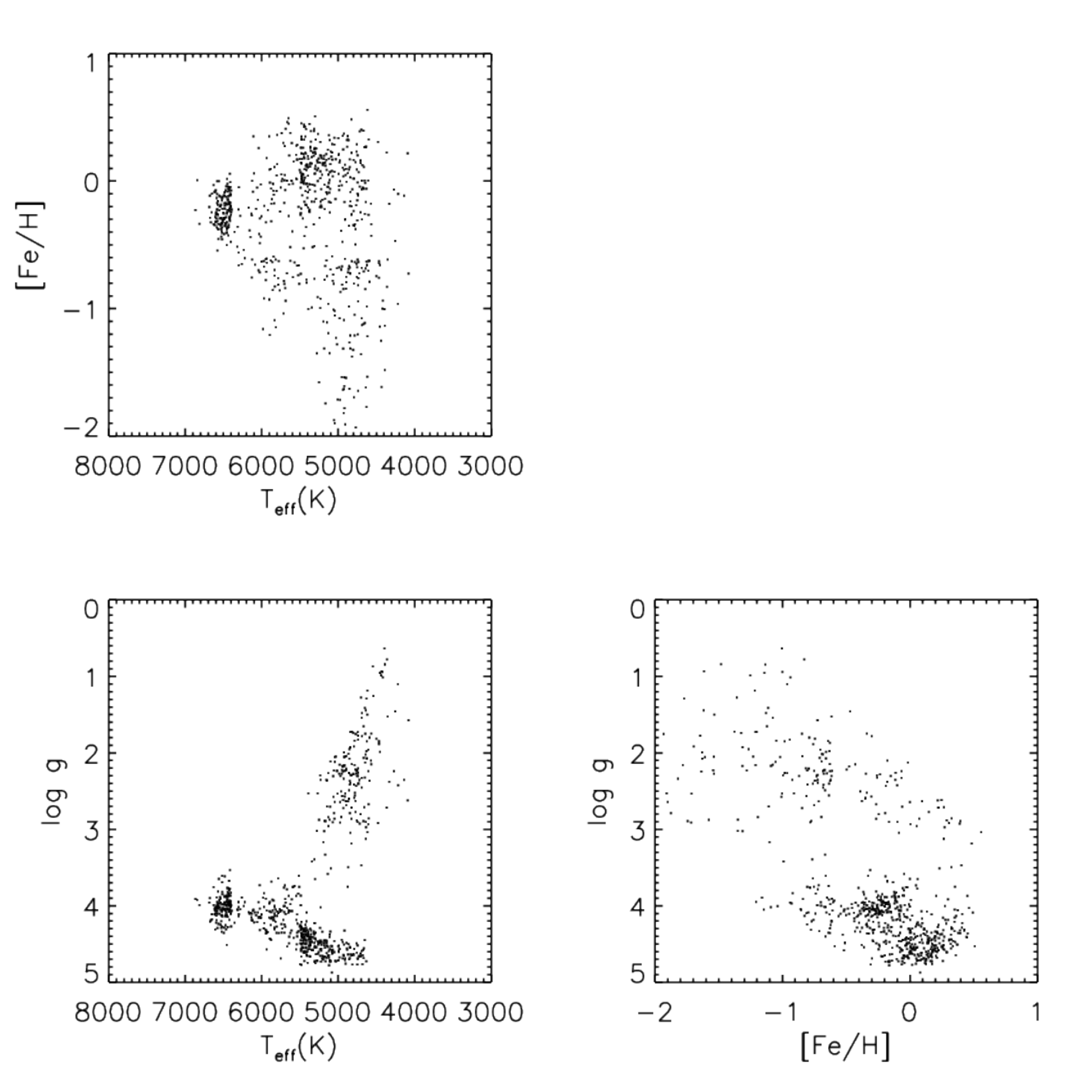}
 \caption[]{The distribution of stellar parameters for our training set. These parameters were taken directly from the GALAH survey \protect\citep{Buder2018}. The training sample was selected to cover a broad range of F, G and early K stars.} 
  \label{trainplot}
 \end{figure*}

As we are making use of {\it Skymapper} data, our method requires a catalogue of stellar parameters for relatively bright southern stars. For this we chose the GALAH sample of \cite{Buder2018}. This provides stellar parameters in the temperature range 4000\,K$<T_{eff}<$7000\,K over a range of surface gravities and metallicities. From this we then selected a training sample which spanned a range of stellar parameters and which were not flagged by \cite{Buder2018} as having unreliable stellar parameters. This latter condition removes most dwarfs in the 4000\,K$<T_{eff}<$4600\,K range, hence we will not be able to determine stellar parameters for dwarfs in this temperature range. We note that this range has relatively few objects in the TESS CTL. This is a result of the selection algorithm of the CTL prioritising bright stars or the smallest, coolest stars with a gap for mid-K dwarfs (see \citealt{Stassun2018}'s Section~4.2.3). Figure~\ref{trainplot} shows all the stars in our training sample. We then applied our photometric quality cuts for each of our nine colours producing training samples for each colour. These samples each comprised around 1000 stars.  

\subsubsection{Dealing with extinction in the training sample}
For our training sample we used the Rayleigh-Jeans Color Excess method proposed by \cite{Majewski2011} and refinded by \cite{Zasowski2013} to correct for reddening. This uses the temperature insensitivity of the $H$-$W2$ colour to estimate reddening. Here we use the same method .

To convert the extinction values from the Rayleigh-Jeans Color Excess method to extinctions in each of our photometric bands we used the reddening relation of \cite{Cardelli1989}. We used the compilation of filter effective wavelengths compiled by Brad Tucker at the Mount Stromlo Observatory \footnote{\url{http://www.mso.anu.edu.au/~brad/filters.html}} for all filters except the Gaia $G$ band. For this last filter we used the effective wavelength calculated by Morgan Fouesneau of MPIA \footnote{\url{http://mfouesneau.github.io/docs/pyphot/}}. 

{To minimise the effect of reddening on our training sample we restricted ourselves to regions of the sky with low extinctions ($A_V<0.05$) in \cite{Schlegel1998}.}

\subsubsection{Our trained model}
\begin{figure*}
 \setlength{\unitlength}{1mm}
 \begin{tabular}{cc}
 \includegraphics[scale=0.27]{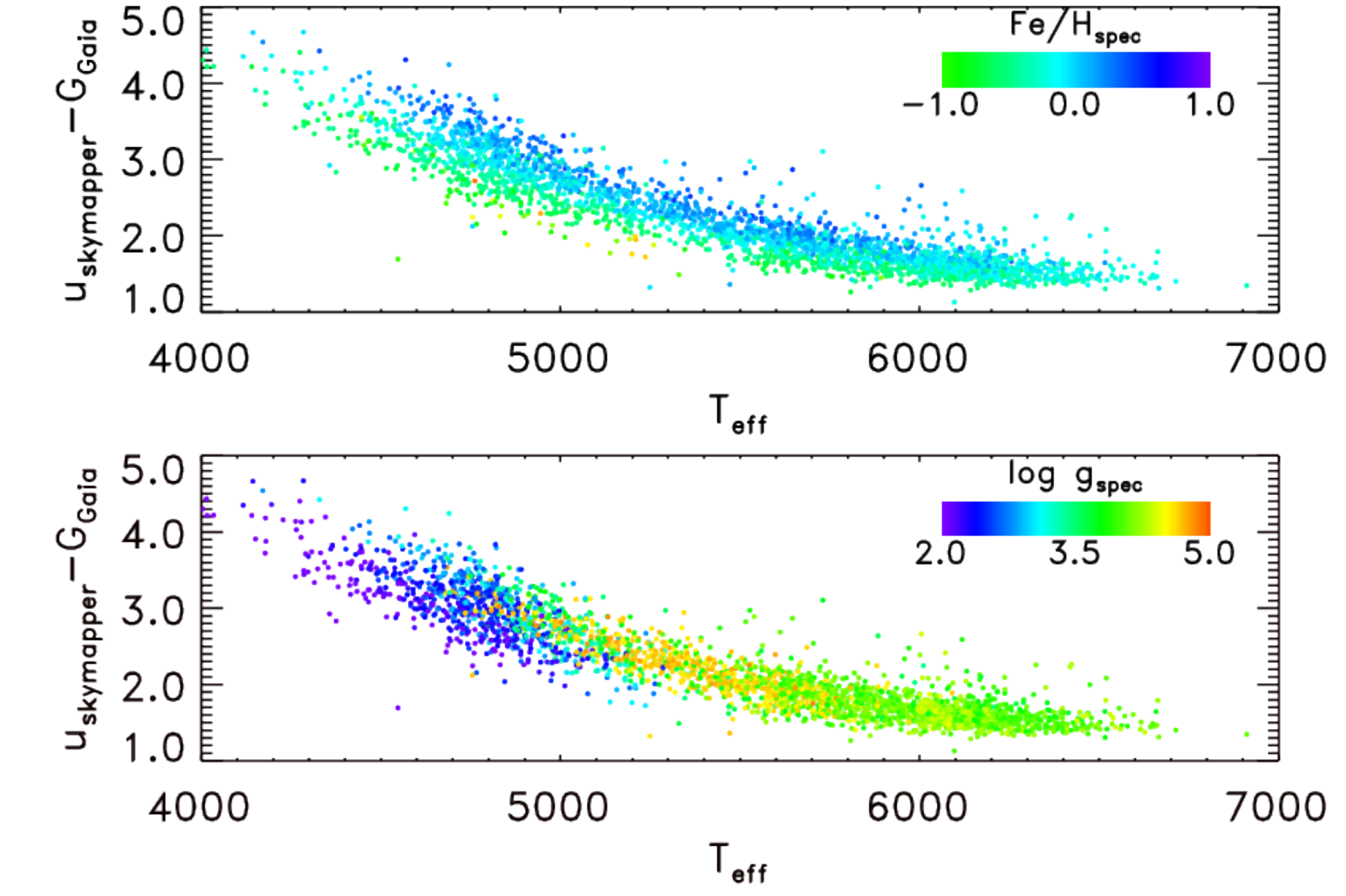}
&
 \includegraphics[scale=0.27]{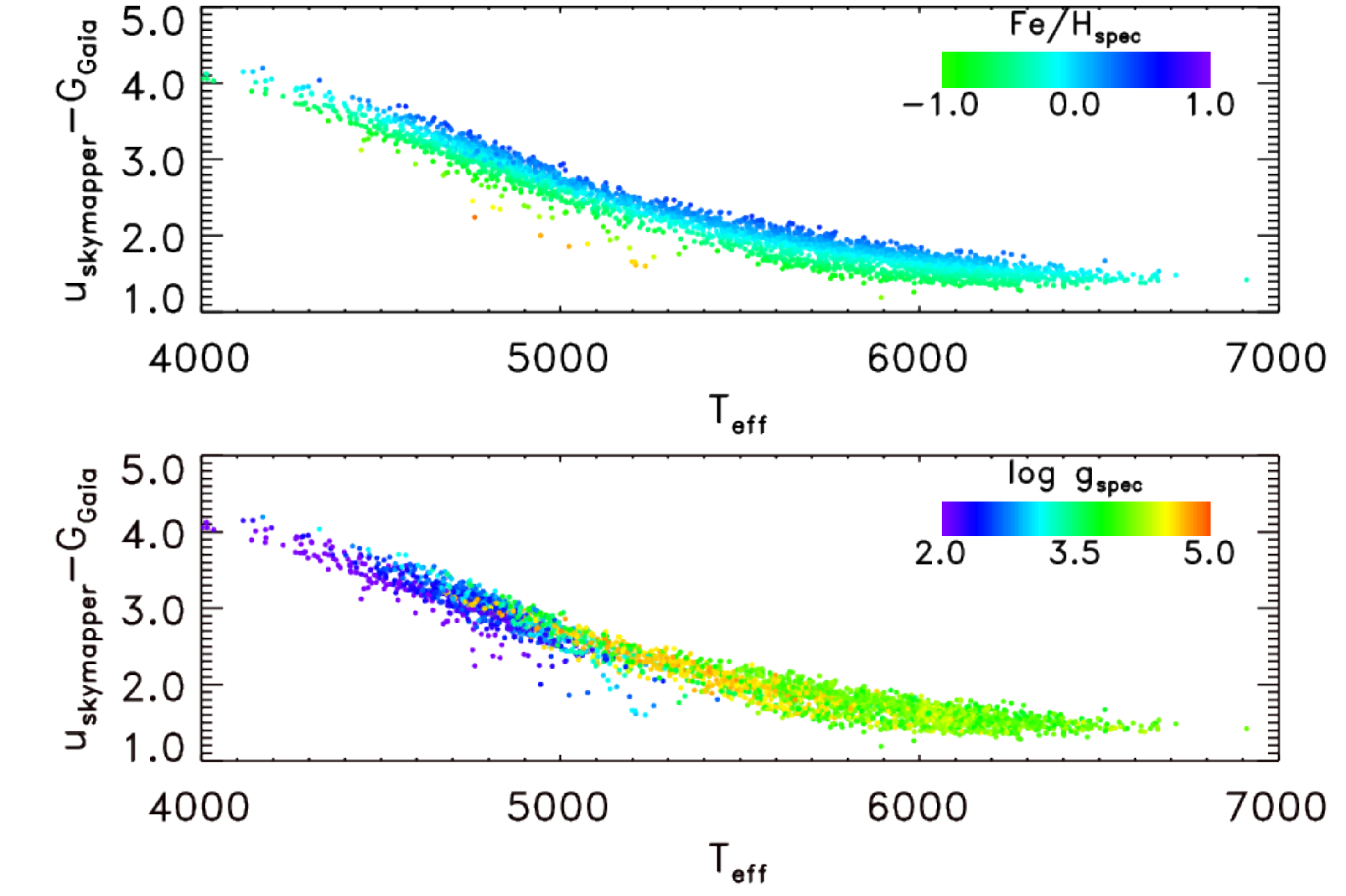}
\\
 
 \includegraphics[scale=0.27]{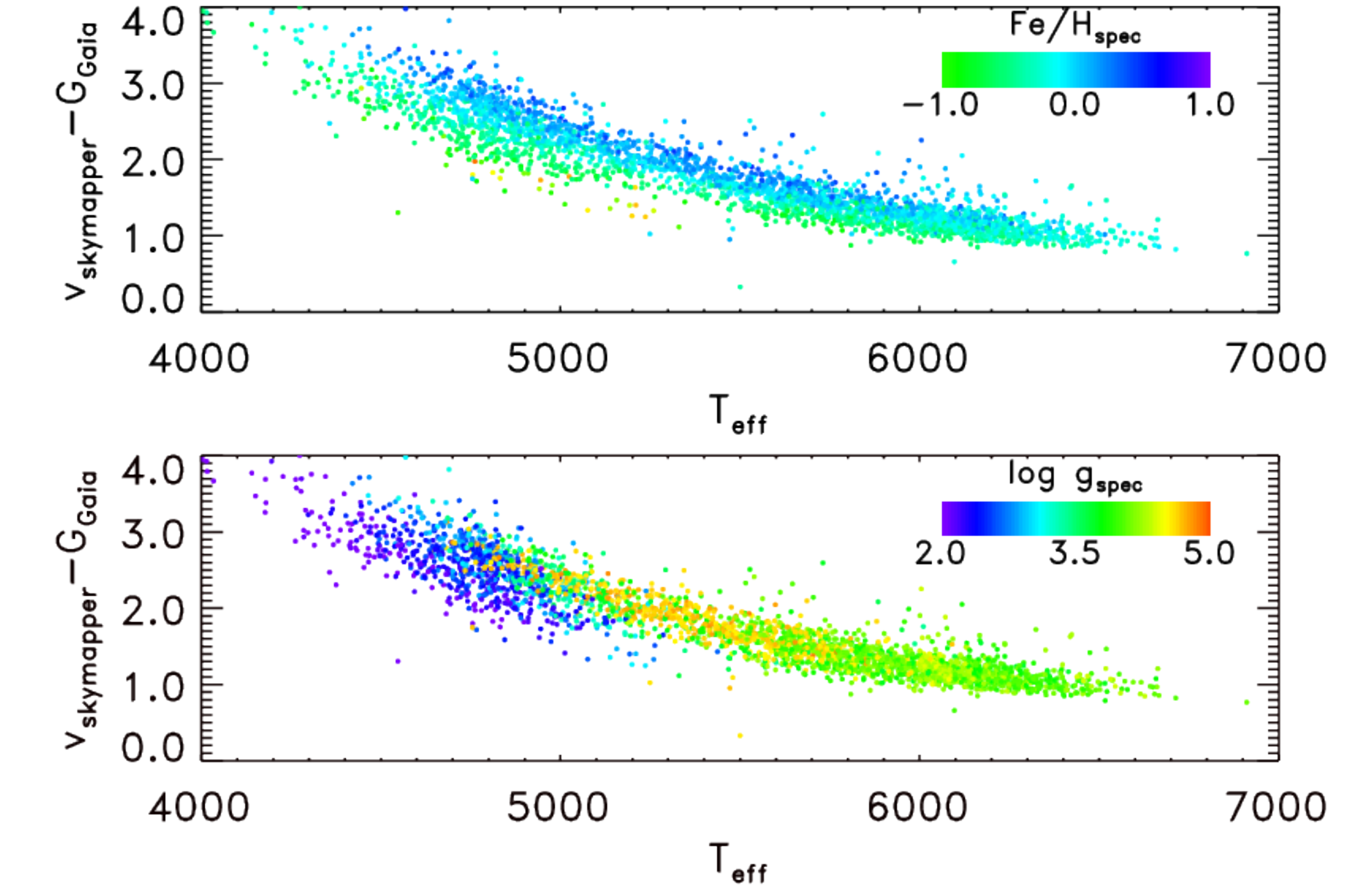}
&
 \includegraphics[scale=0.27]{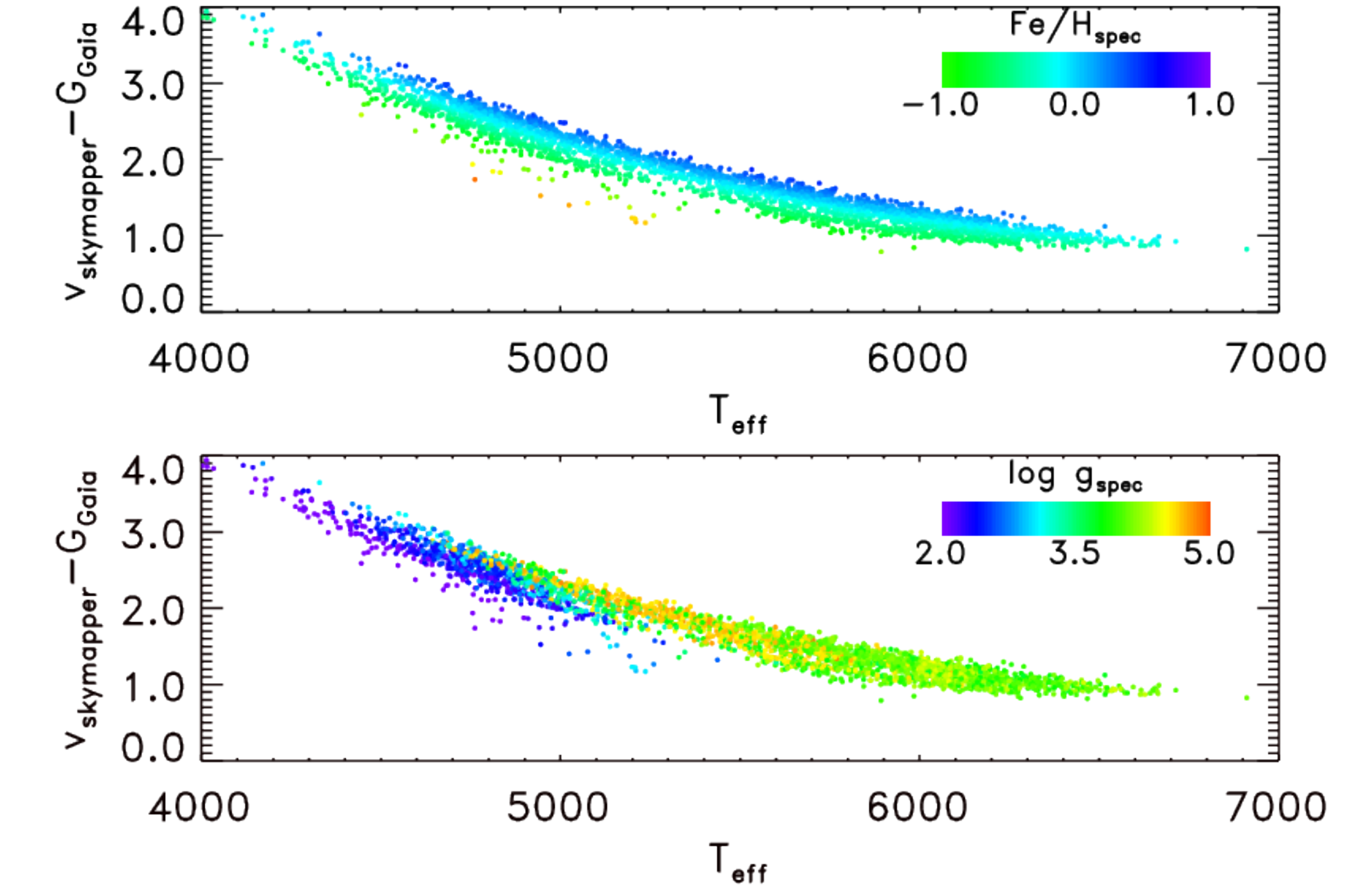}
\\

 \end{tabular}
 \caption[]{Blue optical colours for 2700 objects in the GALAH sample. Left: the objects' observed colours are plotted against their GALAH stellar parameters. Right: the objects' colours are calculated from our trained photometric model and the GALAH stellar parameters. Note how the model reproduces the metallicity sensitivity of both colours. Note we truncate our colour scale at Fe/H=-1. The small number of objects more metal-poor than this will have a similar colour to an Fe/H=-1 object.}
  \label{skymapper_blue_colours}
 \end{figure*}

\begin{figure*}
 \setlength{\unitlength}{1mm}
 \begin{tabular}{cc}
 \includegraphics[scale=0.27]{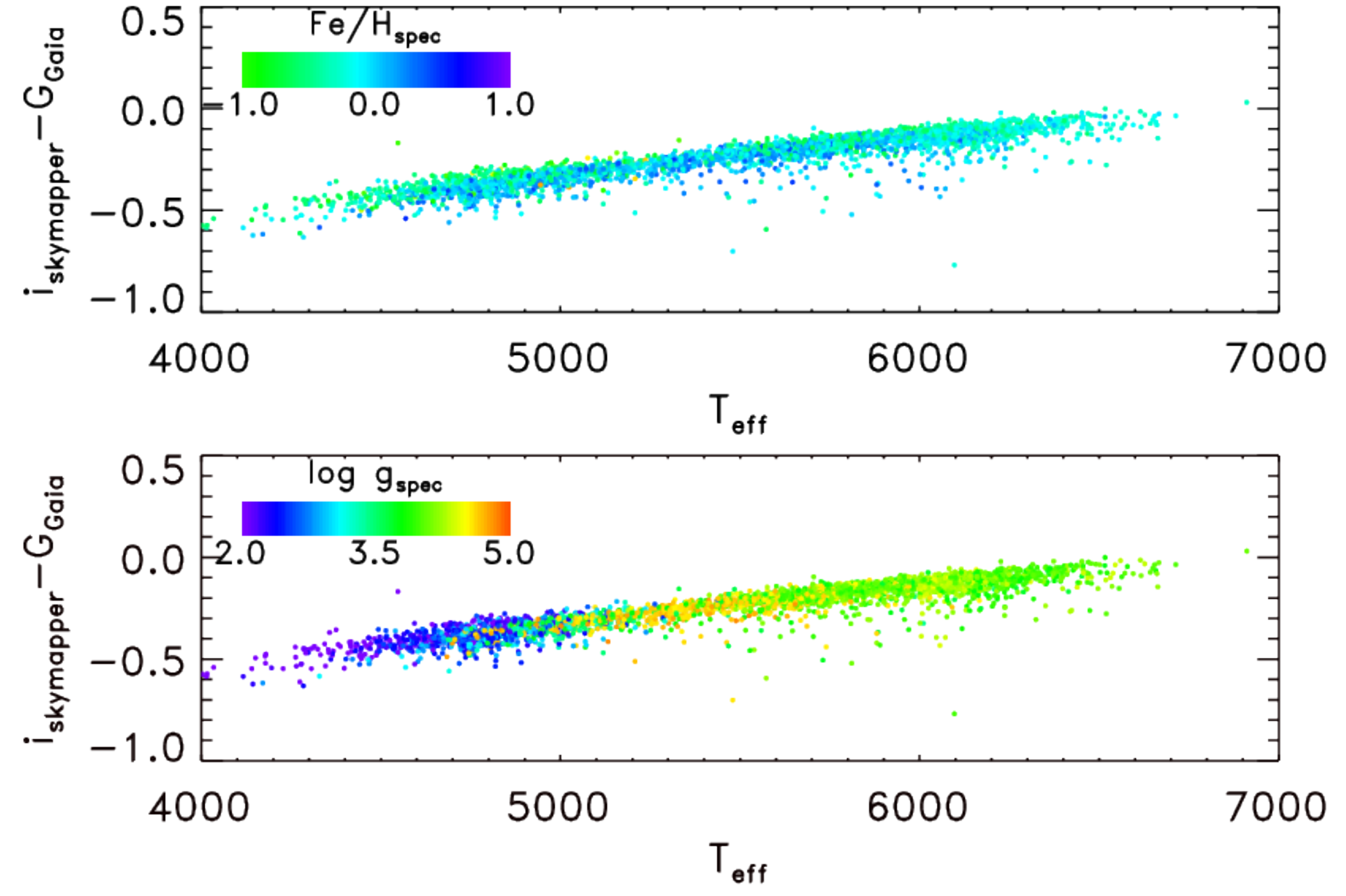}
&
\includegraphics[scale=0.27]{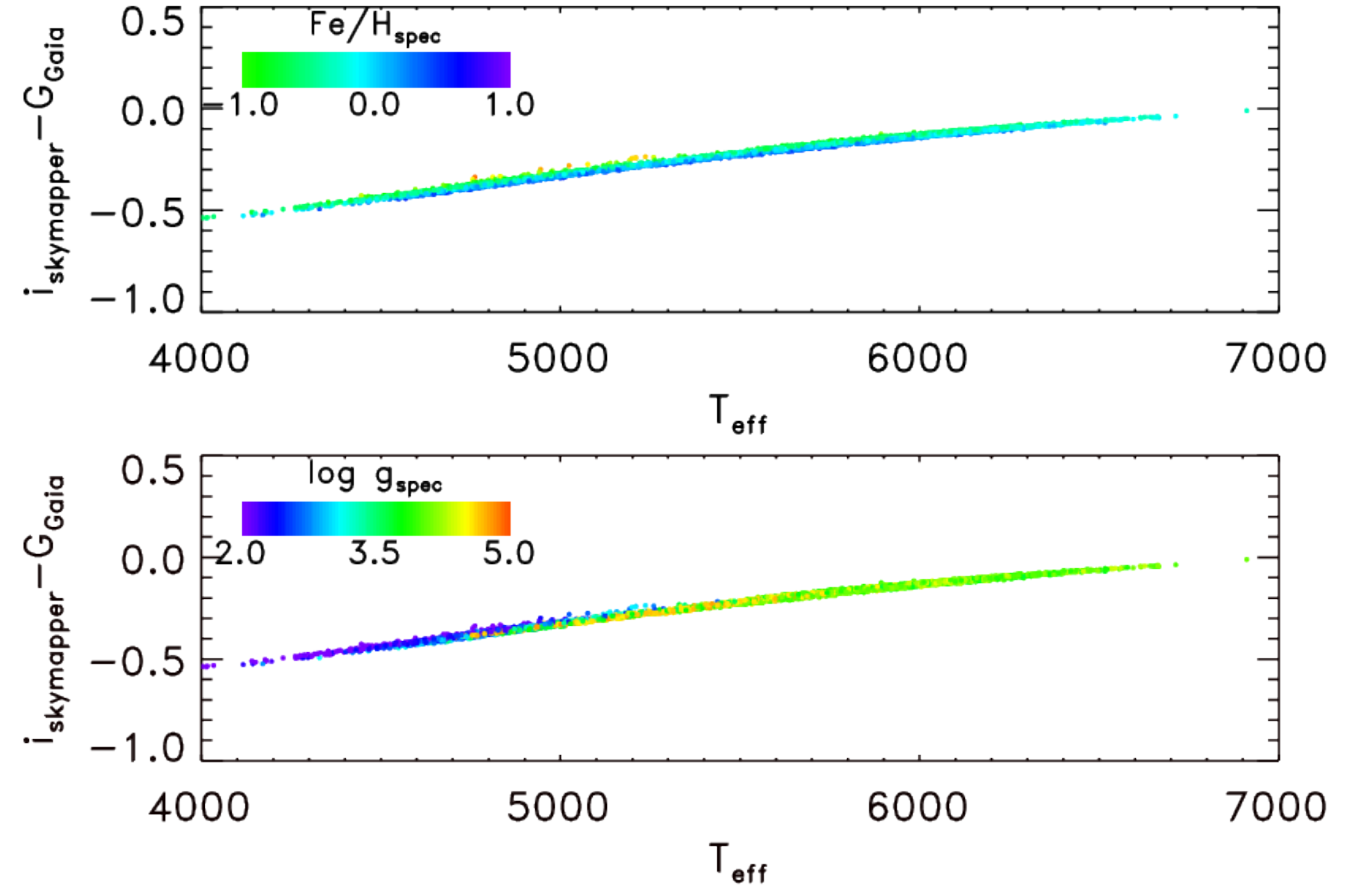}
\\
\includegraphics[scale=0.27]{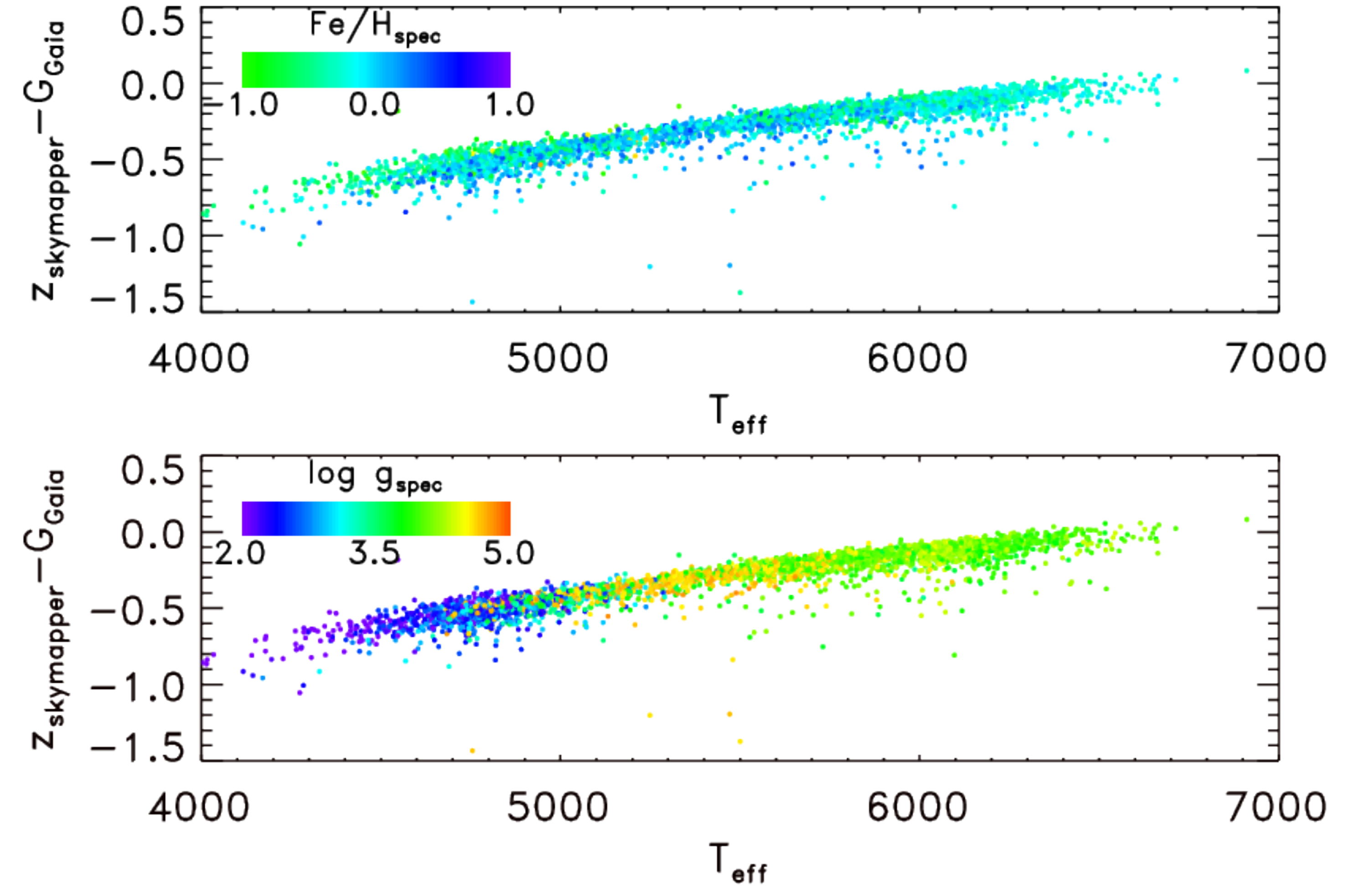}
&
\includegraphics[scale=0.27]{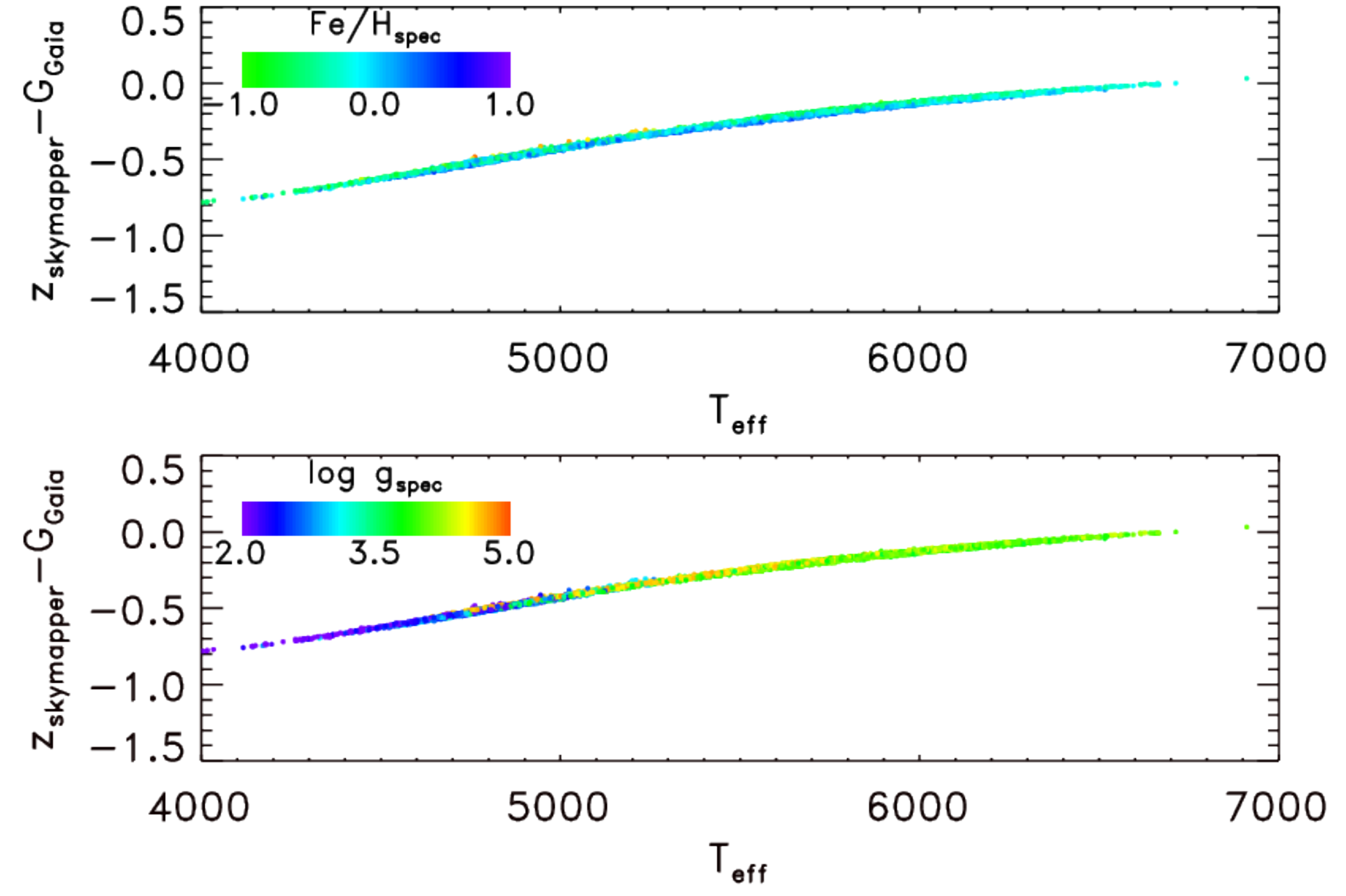}
\\ 
 
 \end{tabular}
 \caption[]{Red optical for 2700 objects in the GALAH sample. Left: the objects' observed colours are plotted against their GALAH stellar parameters. Right: the objects' colours are calculated from our trained photometric model and the GALAH stellar parameters. Note we truncate our colour scale at Fe/H=-1. The small number of objects more metal-poor than this will have a similar colour to an Fe/H=-1 object.} 
  \label{skymapper_red_colours}
 \end{figure*}

\begin{figure*}
 \setlength{\unitlength}{1mm}
 \begin{tabular}{cc}
 \includegraphics[scale=0.27]{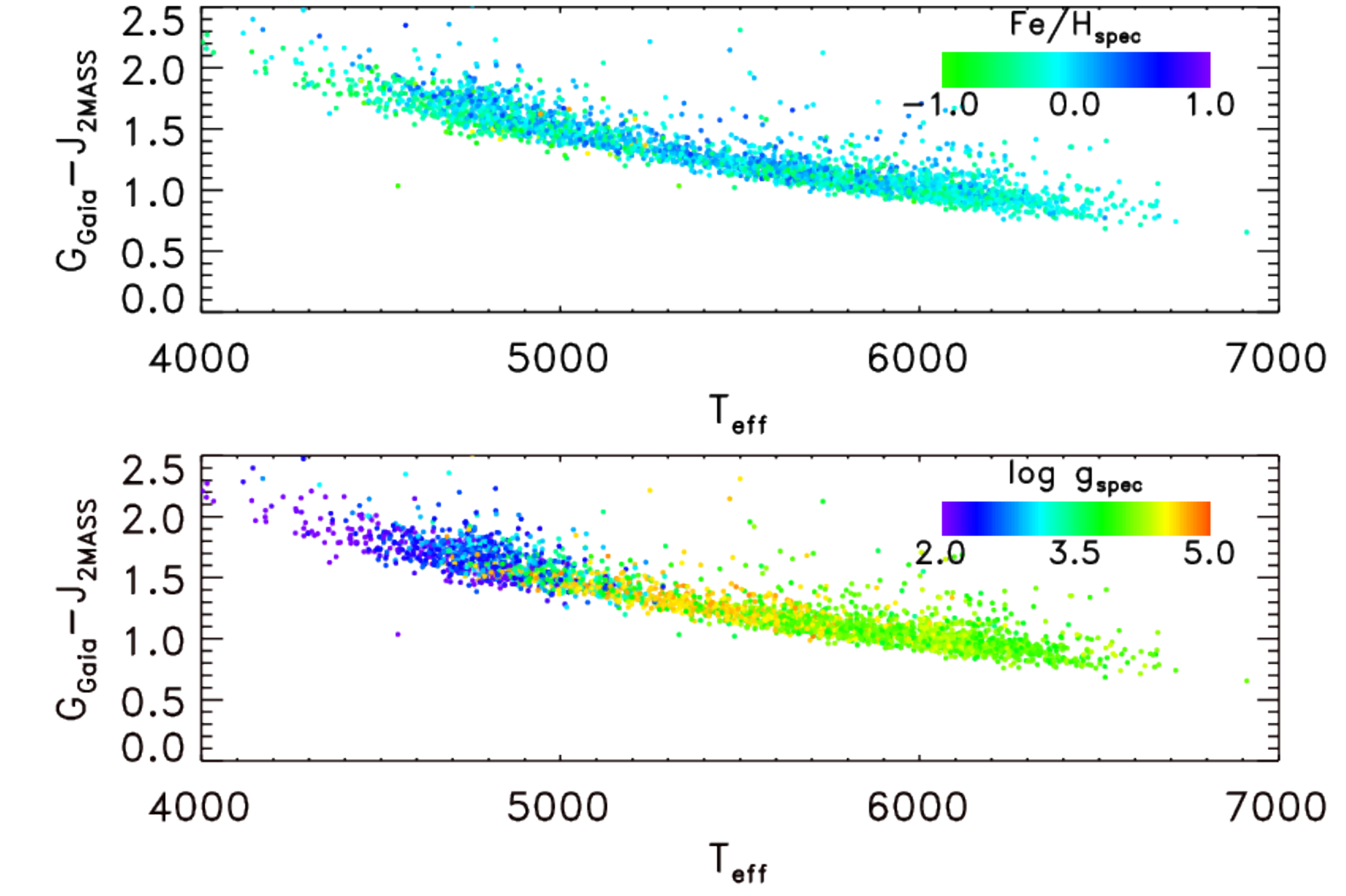}
&
\includegraphics[scale=0.27]{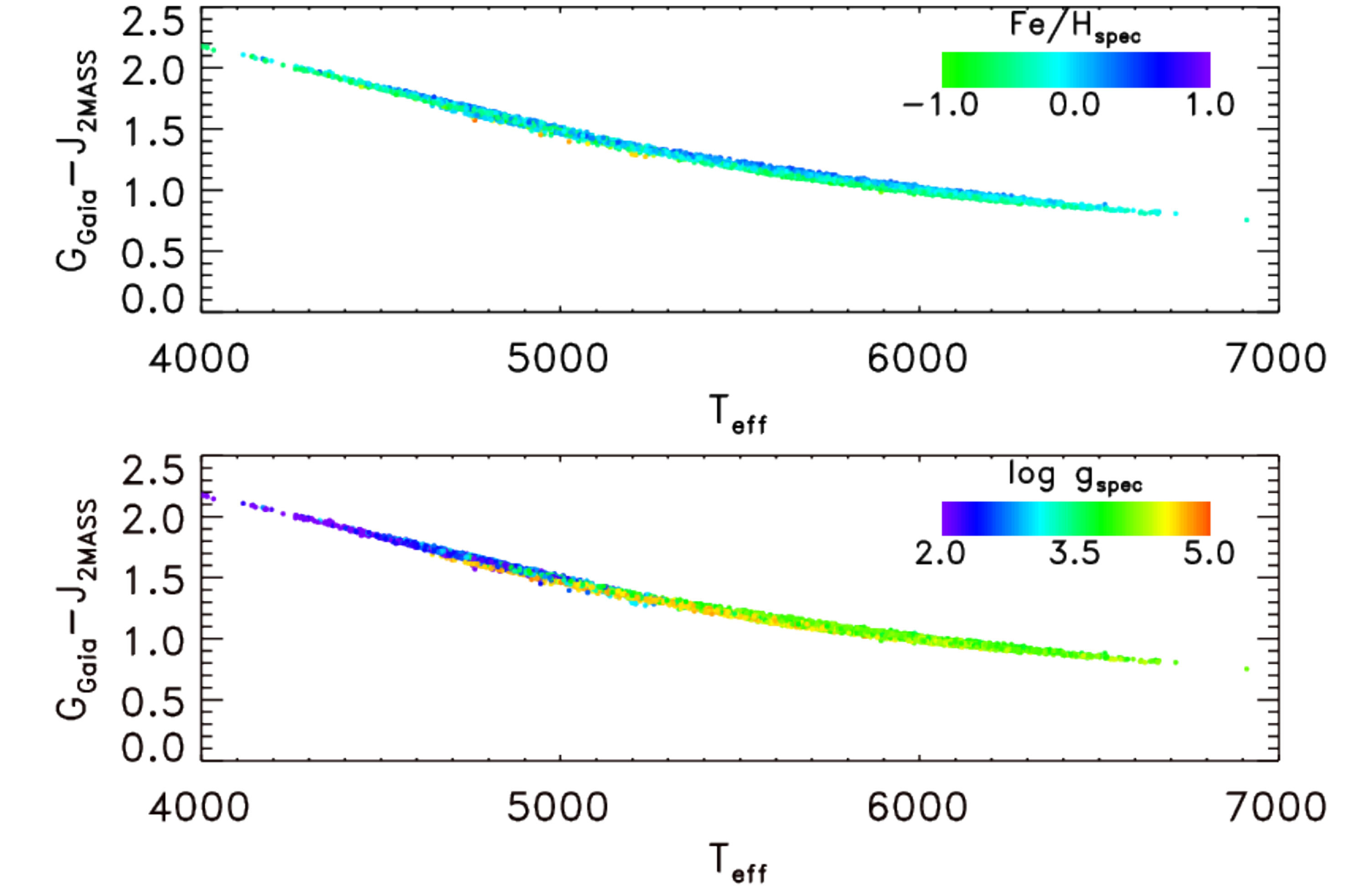}
\\
\includegraphics[scale=0.27]{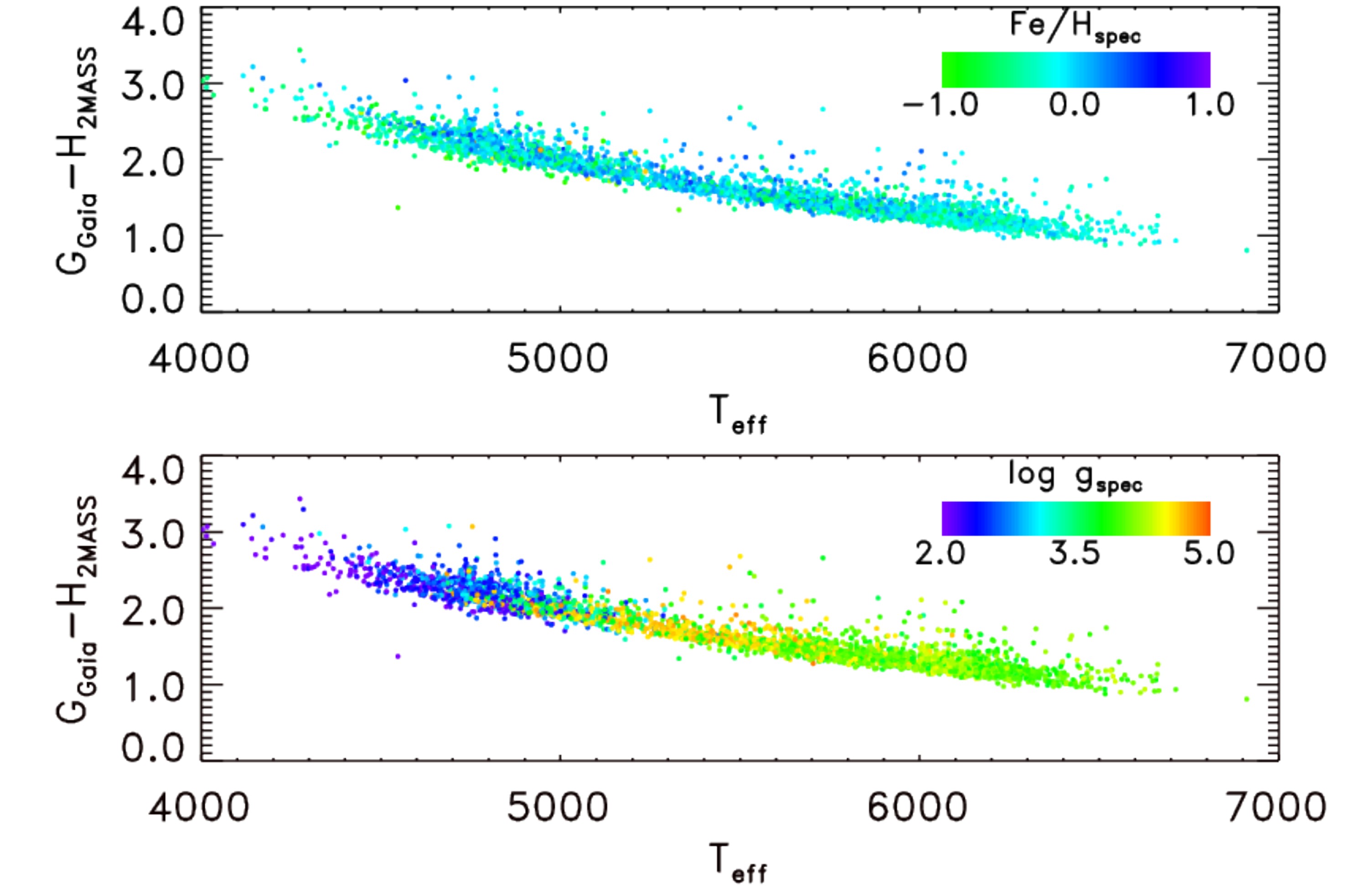}
&
\includegraphics[scale=0.27]{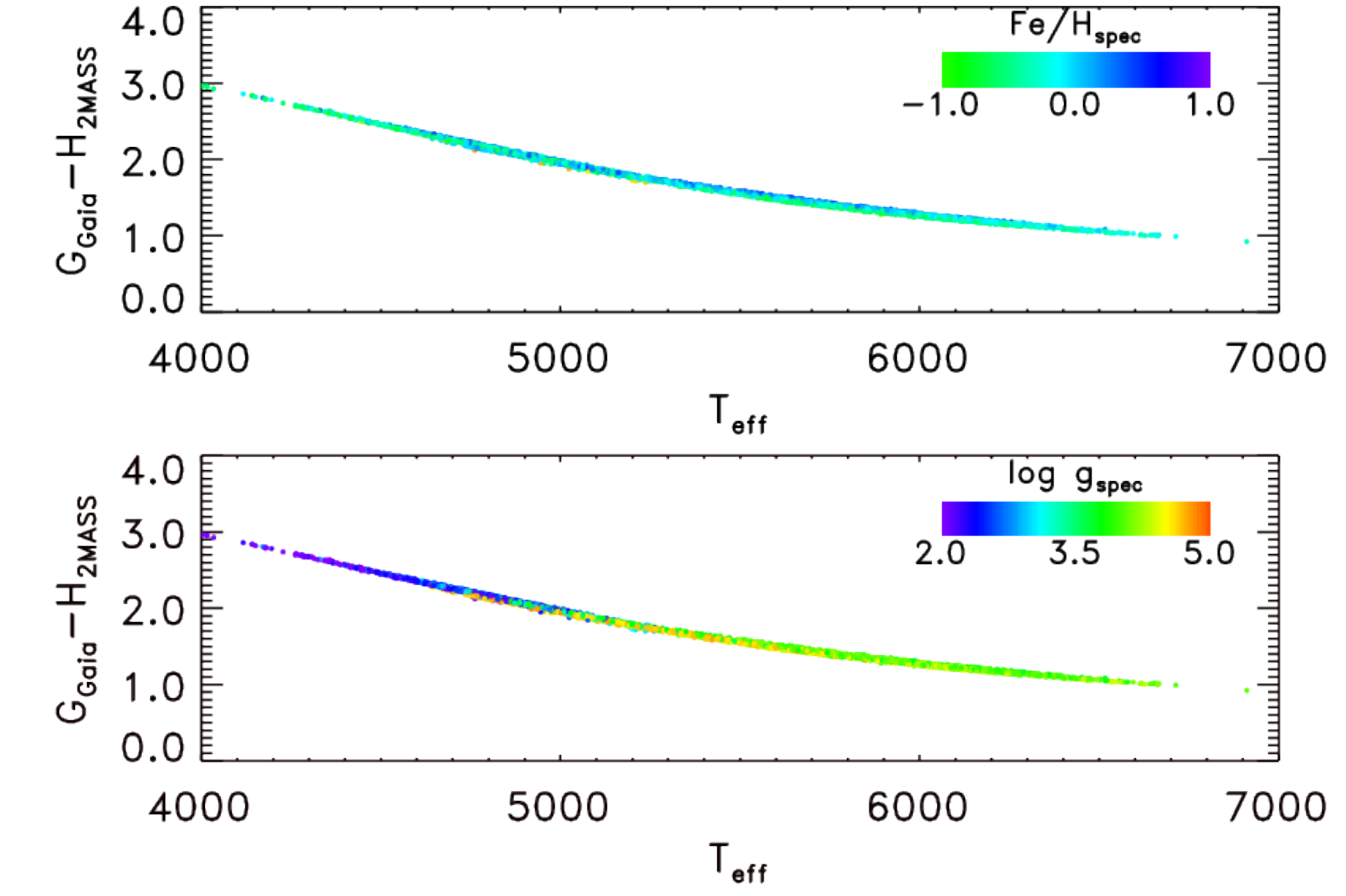}
\\

\includegraphics[scale=0.27]{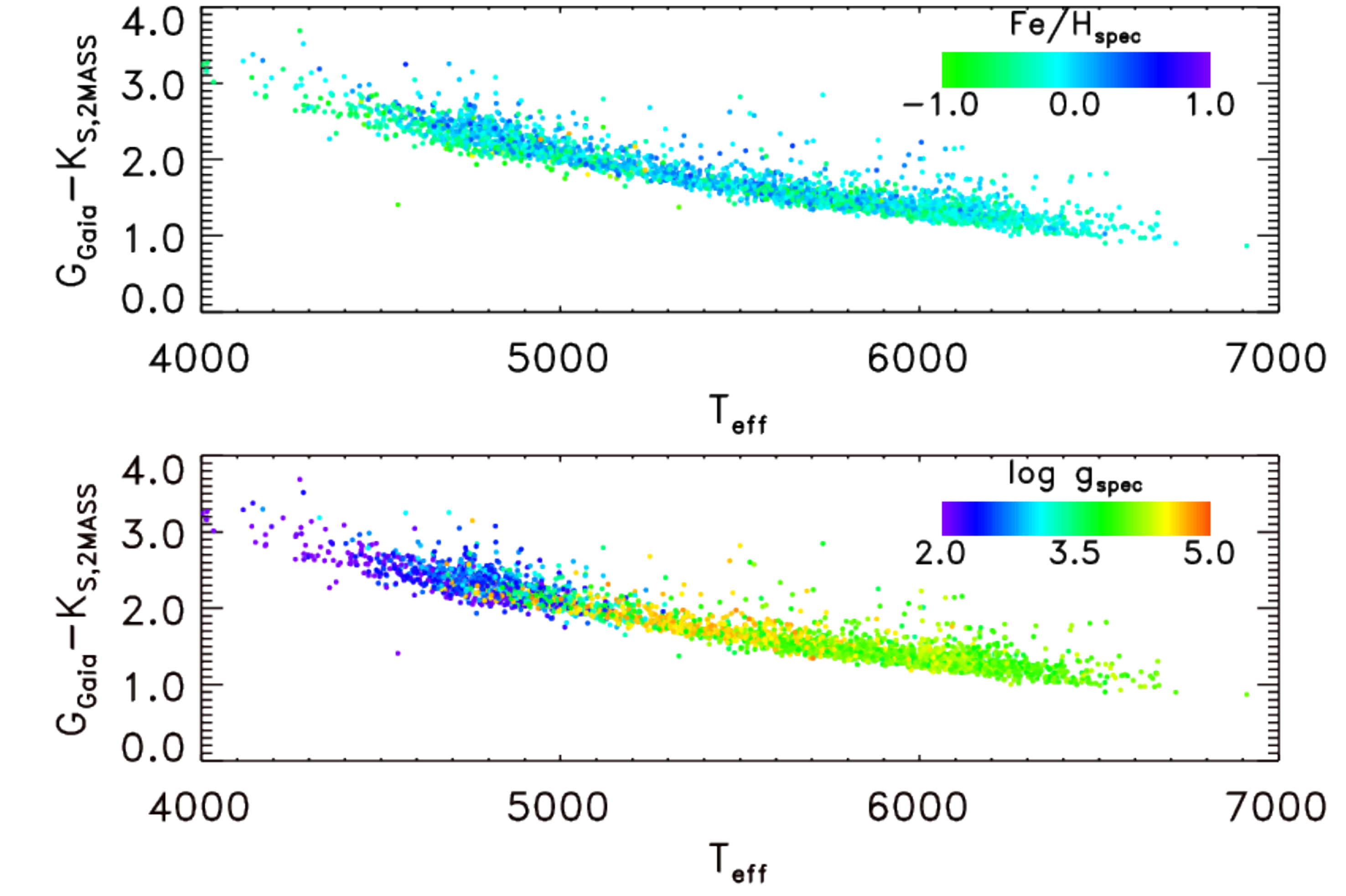}
&
\includegraphics[scale=0.27]{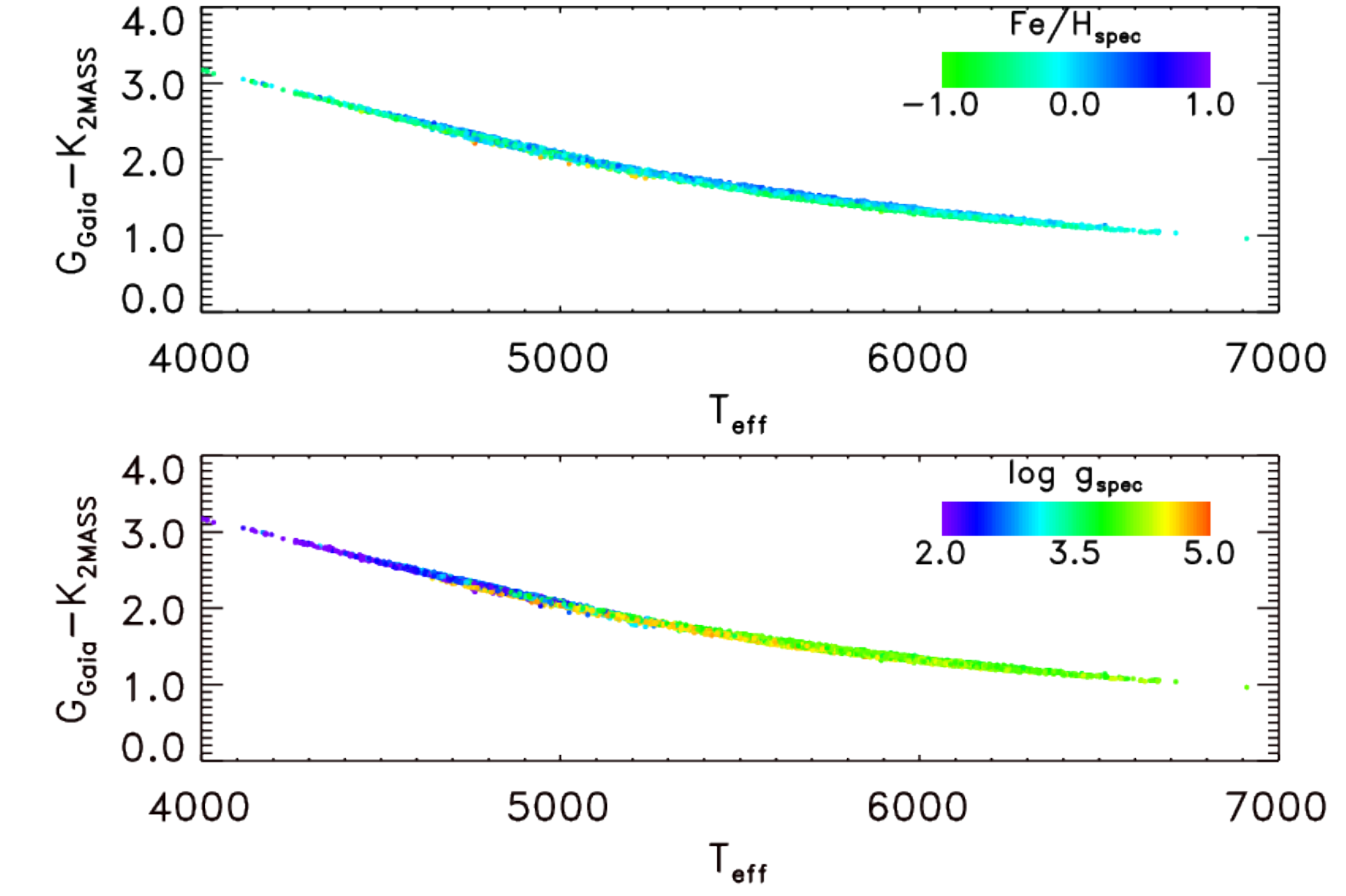}
\\
 \end{tabular}
 \caption[]{{\it 2MASS} near-infrared colours for 2700 objects in the GALAH sample. Left: the objects' observed colours are plotted against their GALAH stellar parameters. Right: the objects' colours are calculated from our trained photometric model and the GALAH stellar parameters. Note we truncate our colour scale at Fe/H=-1. The small number of objects more metal-poor than this will have a similar colour to an Fe/H=-1 object.}
  \label{twomass_colours}
 \end{figure*}

\begin{figure*}
 \setlength{\unitlength}{1mm}
 \begin{tabular}{cc}
 \includegraphics[scale=0.27]{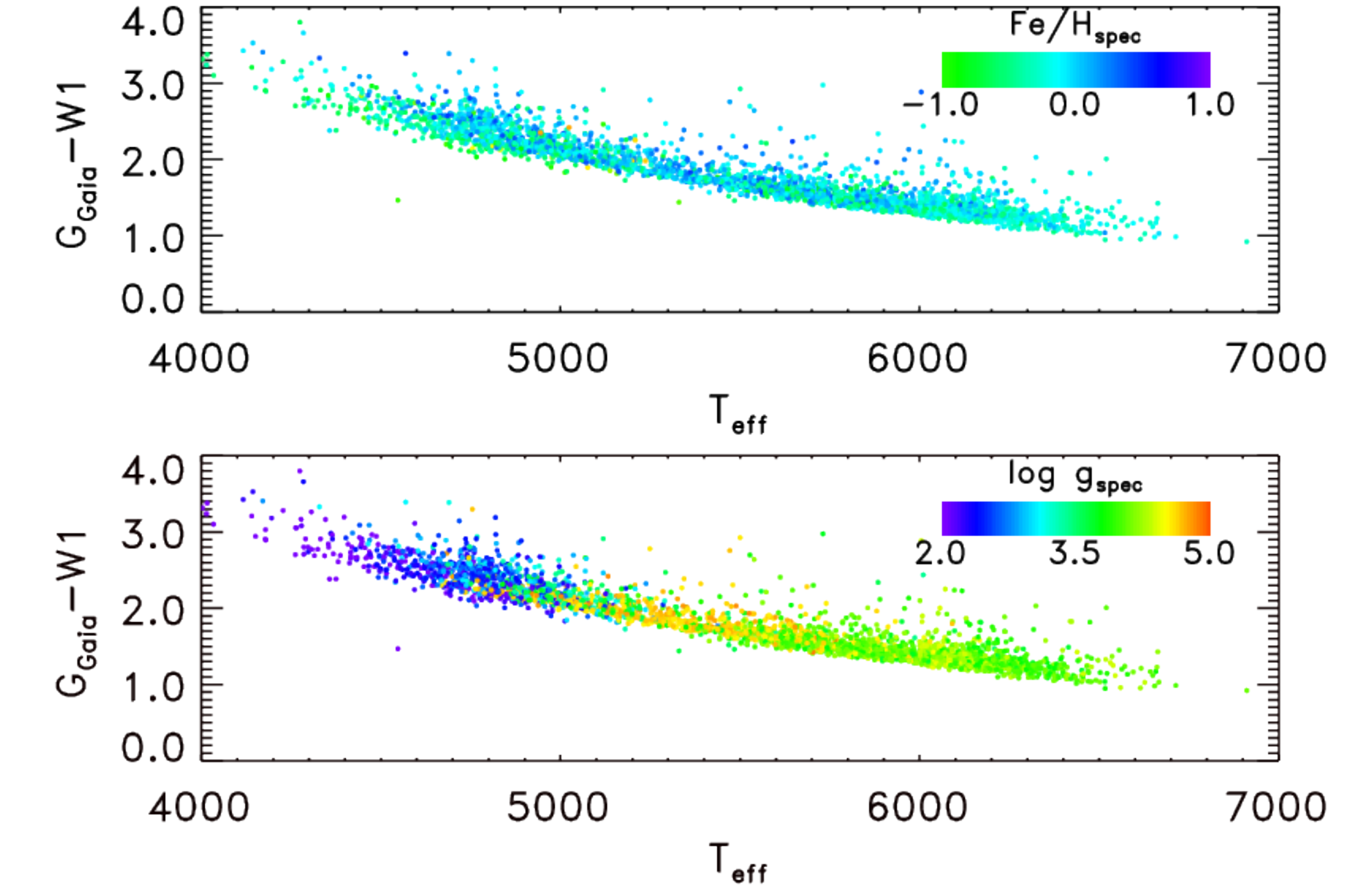}
&
\includegraphics[scale=0.27]{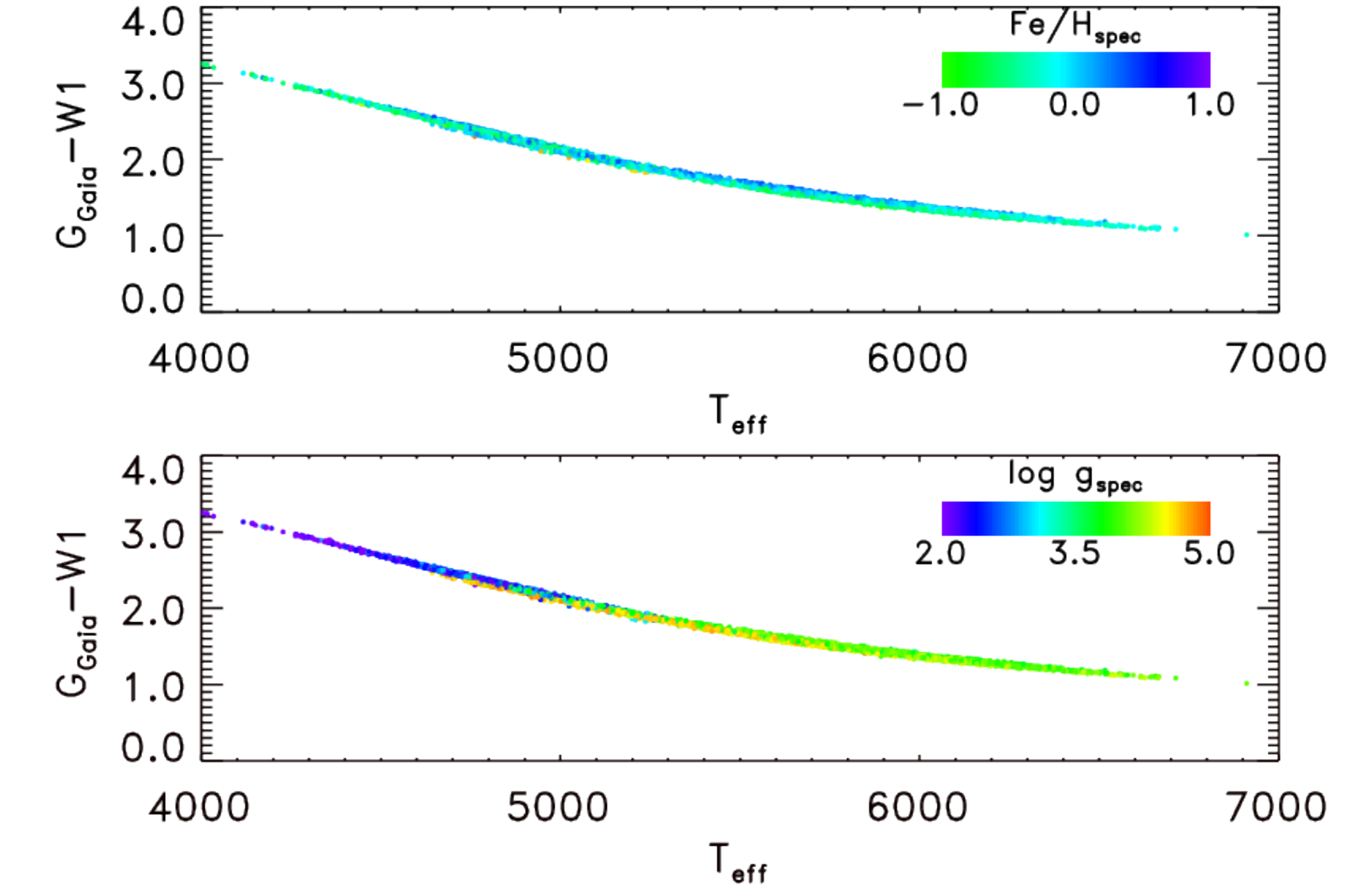}
\\
\includegraphics[scale=0.27]{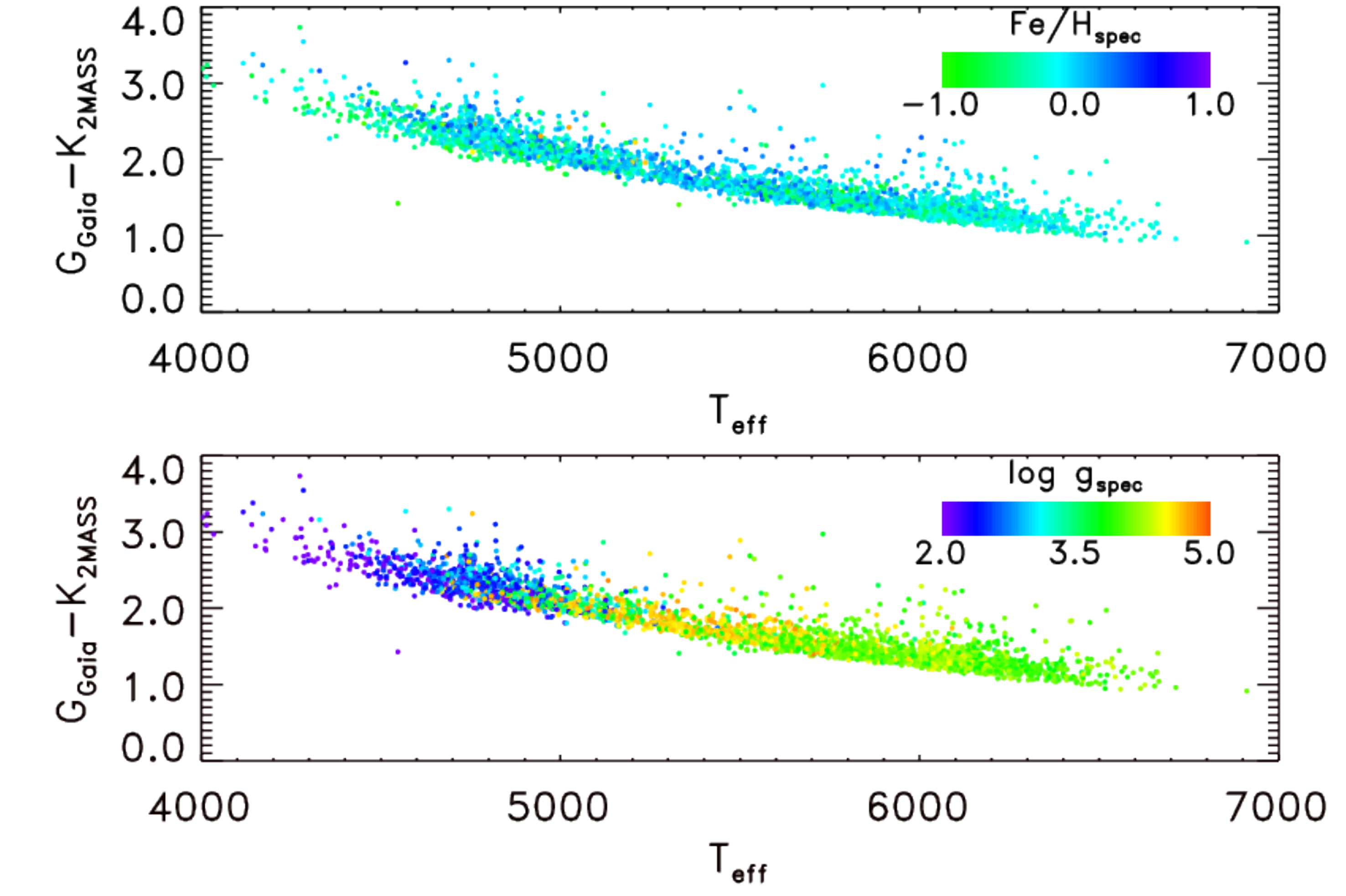}
&
\includegraphics[scale=0.27]{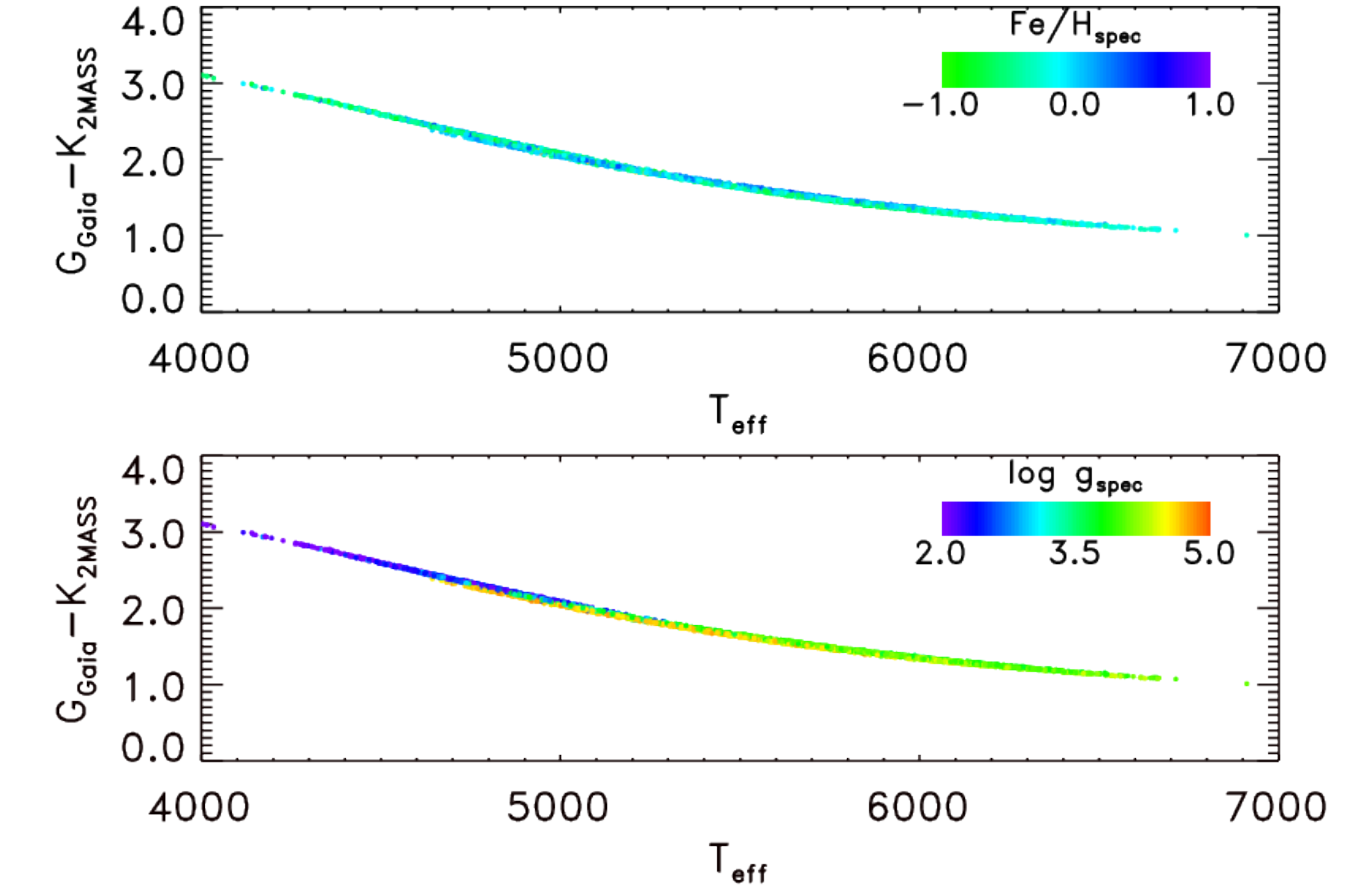}
\\ 
 
 \end{tabular}
 \caption[]{{\it WISE} near-infrared colours for 2700 objects in the GALAH sample. Left: the objects' observed colours are plotted against their GALAH stellar parameters. Right: the objects' colours are calculated from our trained photometric model and the GALAH stellar parameters. Note we truncate our colour scale at Fe/H=-1. The small number of objects more metal-poor than this will have a similar colour to an Fe/H=-1 object.}
  \label{wise_colours}
 \end{figure*}

\begin{figure*}
 \setlength{\unitlength}{1mm}
\includegraphics[scale=0.55]{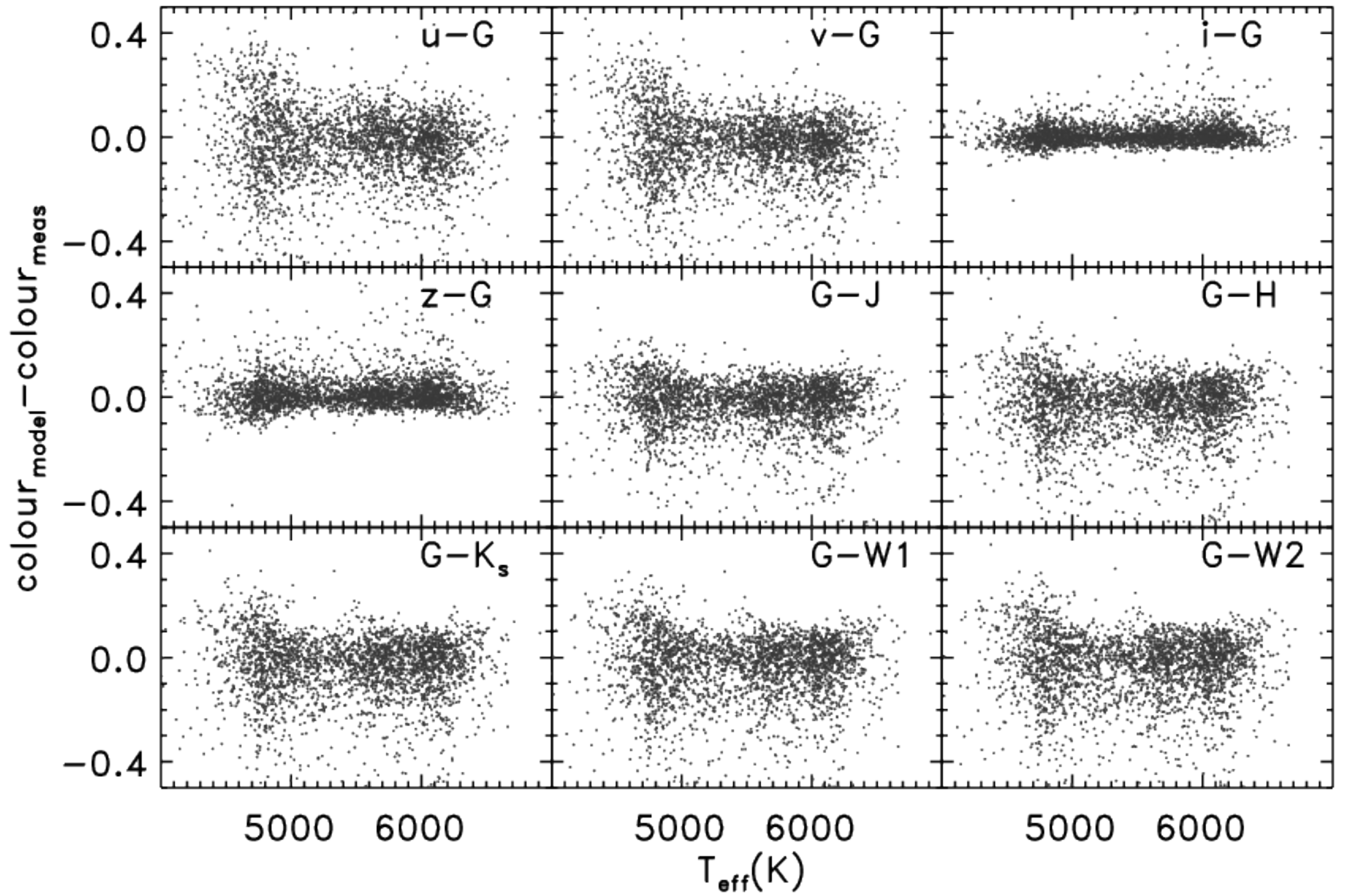}
 \caption[]{A comparison between the colours generated by our photometric model (using the measured GALAH stellar parameters as inputs) and the observed magnitudes for each of for 2700 objects in the GALAH sample. Note how the scatter on each colour difference scales with the scatter estimates in our model (see the last column in Table~\protect\ref{params}).} 
  \label{colour_offset}
 \end{figure*}

For each colour we then ran a simple MCMC to fit the polynomials described in Equation~\ref{model}. We selected the set of parameters from this chain with the highest calculated likelihood. Table~\ref{params} shows these parameter values. Figures~\ref{skymapper_blue_colours},~\ref{skymapper_red_colours},~\ref{twomass_colours}~and~\ref{wise_colours} show how our model reproduces patterns in the data. Figure~\ref{colour_offset} shows the difference between colours generated from our photometric model (using the measured GALAH stellar parameters for each star) and the observed colours. In all cases the residual differences appear as scatters around zero with no clear underlying trend. We note that the larger the measured scatter is in our model (see the last column of Table~\ref{params}) the larger the scatter on the appropriate panel of Figure~\ref{colour_offset} appears. Each of the four aforementioned plots show 2700 objects, selected from GALAH stars not in our training sample to cover a range of effective temperatures, surface gravities and metallicities. The left-hand panels show the objects' GALAH stellar parameters and their observed colours. The right-hand panels show the same objects but with colours calculated from our model using the GALAH stellar parameters. It is clear that our model reproduces the trends in the data relatively well.

\begin{table*}
\caption{The model parameters produced by our GALAH training set. Along with the estimated scatter in absolute $G$ band magnitude calculated from members of the Praesepe cluster.}
\begin{center}
\tiny
\begin{tabular}{lrrrrrrrr}
\hline
Colour&$a_0$&$a_{1}$&$a_{2}$&$a_{3}$&$a_{4}$&$a_{5}$&$a_{6}$&$s_i$(mag.)\\
\hline
$G_{abs,gaia}$&\ldots&\ldots&\ldots&\ldots&\ldots&\ldots&\ldots&0.153\\
$u_{Skymapper}-G_{gaia}$&13.617766&$-$23.443872&$-$11.141706&42.378030&$-$18.535744&$-$0.057528&0.461328&0.137593\\
$v_{Skymapper}-G_{gaia}$&8.099119&$-$19.189484&6.216356&14.295625&$-$7.236286&0.011229&0.440914&0.135398\\
$i_{Skymapper}-G_{gaia}$&0.339292&$-$0.348567&0.800790&$-$2.106884&0.987482&0.001888&$-$0.025408&0.047475\\
$z_{Skymapper}-G_{gaia}$&$-$0.151847&0.947015&0.955333&$-$4.218476&2.008298&0.011817&$-$0.026462&0.067180\\
$G_{gaia}-J_{2MASS}$&0.169107&2.514782&$-$7.450069&10.046609&$-$3.710285&$-$0.029018&0.055635&0.095357\\
$G_{gaia}-H_{2MASS}$&1.055295&$-$0.535033&$-$4.584718&10.398408&$-$4.310333&$-$0.025798&0.048321&0.117178\\
$G_{gaia}-K_{s,2MASS}$&0.753974&0.380586&$-$5.060822&10.095362&$-$4.01346&$-$0.035459&0.068763&0.124873\\
$G_{gaia}-W1_{WISE}$&2.646238&$-$3.817434&$-$4.161496&13.381048&$-$5.820988&$-$0.036533&0.064117&0.132975\\
$G_{gaia}-W2_{WISE}$&1.686011&$-$0.386909&$-$8.792007&16.017826&$-$6.395953&$-$0.027864&0.032205&0.137357\\
\hline

\end{tabular}
\normalsize
\end{center}
\label{params}
\end{table*}



\subsection{Markov Chain Monte Carlo parameter estimation}
We have built models of how stellar colours relate to effective temperature, gravity and metallicity. We now use these as part of a MCMC effort to explore our likelihood space and measure the parameters of each star. 

We begin by taking a grid of PARSEC stellar models and interpolating them so that they form a regular grid in $\log$~mass, initial metallicity and $\log$~age. For each object we begin with a 1\,Gyr-old, 2\,$M_{\odot}$ star with [Fe/H]=$-$0.1 and $A_V=0.001$\,mag. For an initial 10,000 steps we only allow the mass to vary, allowing each star to find its approximate stellar evolution stage. We then step around on our grid of mass, metallicity and age; also stepping around in extinction. At each step we generate an effective temperature, $\log g$ and absolute Gaia $G$ magnitudes from the PARSEC model grid. We then used the effective temperature, $\log g$ and metallicity as inputs for our trained photometric model, yielding estimates of our nine colours. These colours then had extinction corrections applied to them calculated from the value of $A_V$ our model found itself at and the relationship between $A_V$ and the extinction in other filters that we previously calculated. The colours, with extinction included, were combined with the model-derived absolute magnitude and along with the observed magnitudes and colours, fed into our likelihood for that star (see Equation~\ref{like3}). We then used a simple Metropolis-Hastings algorithm to compare our likelihood value with the previous step and to decide if we should stay with the step in parameter space we just made. 

We find that this method works well for dwarf stars producing an acceptance rate in the 20--30\% range using initial step sizes of initial step sizes of 0.4\,dex in metallicity, 0.2\,M$_{\odot}$ in mass and 0.2\,Gyr in age. In order to fully explore our likelihood we aim for an acceptance rate in the 20--30\% range. If after our initial burn-in period of 100,000 iterations we find that the acceptance rate has fallen below 20\% we reduce our step sizes by a factor of 1.1. If we find that the acceptance rate goes above 30\%, we increase each step size by a factor of 1.1. We repeat this process until the acceptance rate is acceptable. We run the code for a burn-in period of five hundred thousand steps followed by a further one million steps.  We find that the likelihoods of dwarf stars can be explored with relatively large step sizes. Evolved stars of a particular temperature and brightness follow a narrow channel in mass-age-metallicity space and thus require smaller step sizes.

We set a requirement that they do not exceed the boundaries of our grid ($0.15\,M_{\odot}<$mass$<25\,M_{\odot}$), 0.008\,Gyr$<$13\,Gyr, $-1.85$\,dex$<$[Fe/H]$<$0.65\,dex) and that the effective temperature stays between 3800\,K and 7800\,K (a slightly wider temperature range than our training sample). Our extinction $A_V$ has a requirement that it must be less than the extinction from the integrated Galactic reddening maps of  \cite{Schlegel1998} (generated from the \texttt{dust\_getval} IDL routine) and also less than the extinction upper limit in the Gaia stellar parameters catalogue of \cite{Andrae2018}. We note that this latter source also uses the Gaia $G$ magnitude in its calculation so is not an entirely independent constraint. We found that we were not able to constrain the metallicity of objects which did not have data in one of the $u_{skymapper}$ or $v_{skymapper}$ bands. Hence for such objects we set the metallicity to be -0.1\,dex, the modal value of the solar neighbourhood distribution found by \cite{Gray2006}.

The values we quote for each parameter are for the maximum likelihood solution that our chain found. We also quote 1$\sigma$ confidence limits around these values calculated from the likelihood contours in our chain.
\section{Results}
We ran our parameter estimation code on 1,086,223 objects in the TESS CTL (version 0701). These objects have a) CTL effective temperatures of 4000\,K to 8000\,K, b) coordinates located in the southern celestial hemisphere (where we have Skymapper data), c) Gaia astrometric solutions that have a Reduced Unit-Weighted Error less than 1.5 \citep{Lindegren2018,Lindegren2018a} (this cut removed 11\% of objects that passed all other cuts) and parallaxes more significant than 5$\sigma$ and d) have reliable photometry in at least five of the Skymapper, 2MASS and WISE bands (this removed 6\% of objects that passes all other cuts). This results in a list of likely FGK stars with good Gaia astrometric solutions which are likely to be TESS short cadence targets. Our training set does not contain a significant number of cool dwarfs ($T_{eff}<4600$\,K and $\log g>4$) or hot objects ($T_{eff}>7000$\,K). Hence for objects where we calculate stellar parameters in these ranges we simply quote that they are dwarfs cooler than 4600,K or objects hotter than 7000\,K in our final catalogue. We exclude these objects from our subsequent plots. There are 939,457 objects for which we quote effective temperature and gravity values and 638,972 for which we also quote a metallicity.

The typical effective temperature uncertainty in our sample is 247\,K, slightly higher than the typical 178\,K uncertainty in the TESS CTL for the objects we sampled. Our typical metallicity uncertainty is 0.28\,dex. The higher values on these uncertainties are likely due to our TESS sample having stars which lack data in some of the Skymapper bands. 

\begin{figure*}
 \setlength{\unitlength}{1mm}

\includegraphics[scale=0.28]{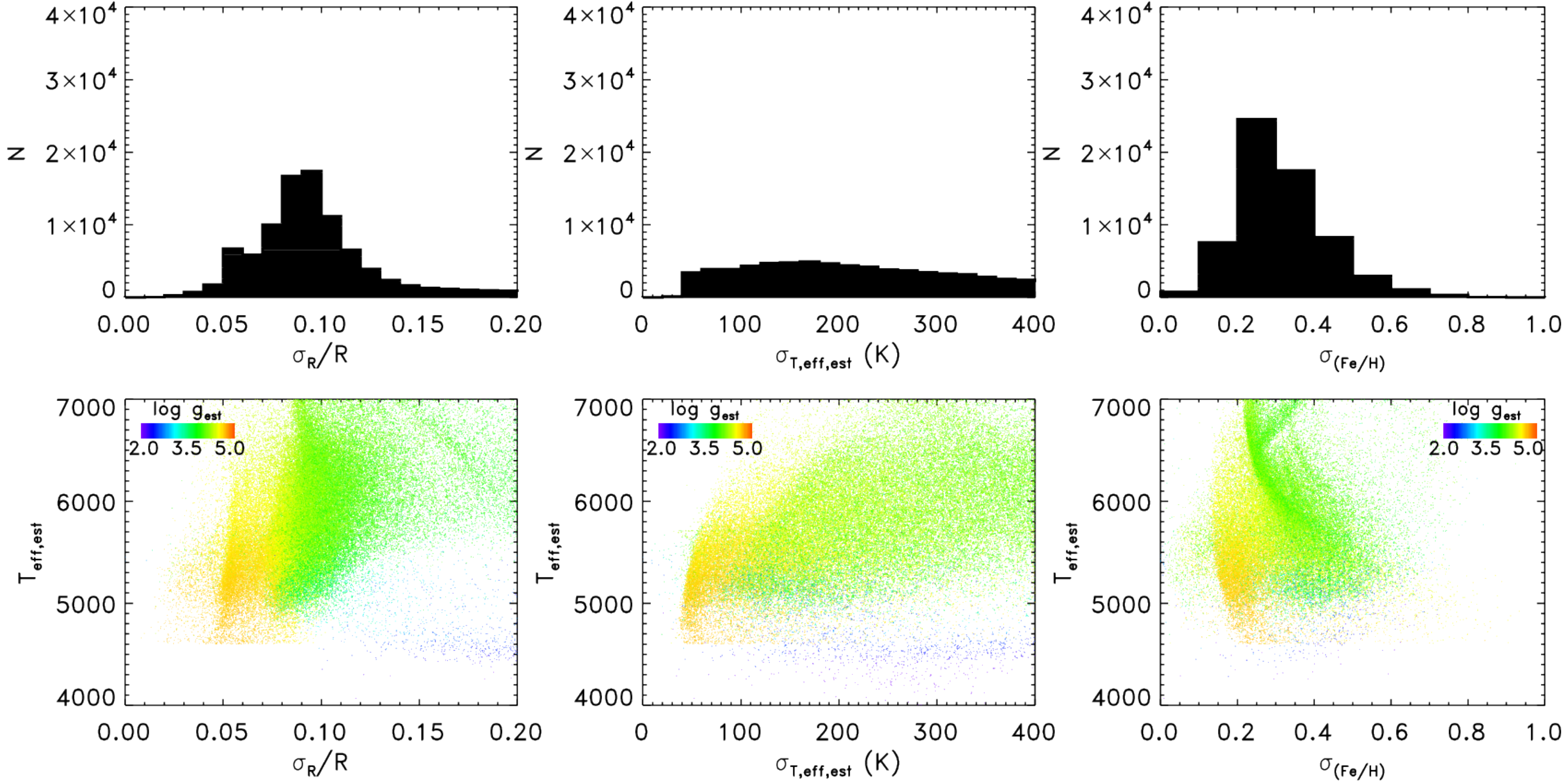}
 \caption[]{Histograms of the uncertainties on the measured stellar parameters for our sample. As we quote separate upper and lower uncertainty bounds for each object we take and average of these to values to obtain an estimate of each objects typical uncertainty. In all three cases cooler objects have lower parameter uncertainties. The range of uncertainties at a given effective temperature likely relates to some objects lacking photometric measurements in some bands. We achieve a typical fractional radius uncertainty of 10\%.  Note the right-hand plots excludes stars which we could not determine the metallicity for as they had no blue Skymapper photometry.}
  \label{err_hist}
 \end{figure*}

Figure~\ref{err_hist} shows histograms on our measured parameters for 95,356 randomly selected stars. Here we take an average of our upper and lower uncertainty estimates to represent the uncertainty for each object. The effective temperature uncertainty is lower for objects cooler than $T_{eff}=6000$\,K while hotter objects have higher uncertainties. This is to be expected as colour changes more rapidly with effective temperature at cooler temperatures. We note that hotter stars often have very high upper temperature bounds due to this effect. The same pattern are seen in the uncertainties in Fe/H and the
fractional uncertainties of our $\log R/R_{\odot}$ estimates with hotter stars having higher uncertainties. This latter histogram peaks at 9-10\% with dwarfs having lower uncertainties than giants. Our median radius uncertainty is 9.3\%, reducing to 8.3\% for higher priority objects.

 \clearpage
\subsection{Stellar parameter accuracy}
\subsubsection{Testing with GALAH data}
\begin{figure*}
 \setlength{\unitlength}{1mm}
 \begin{tabular}{cc}
\includegraphics[scale=0.27]{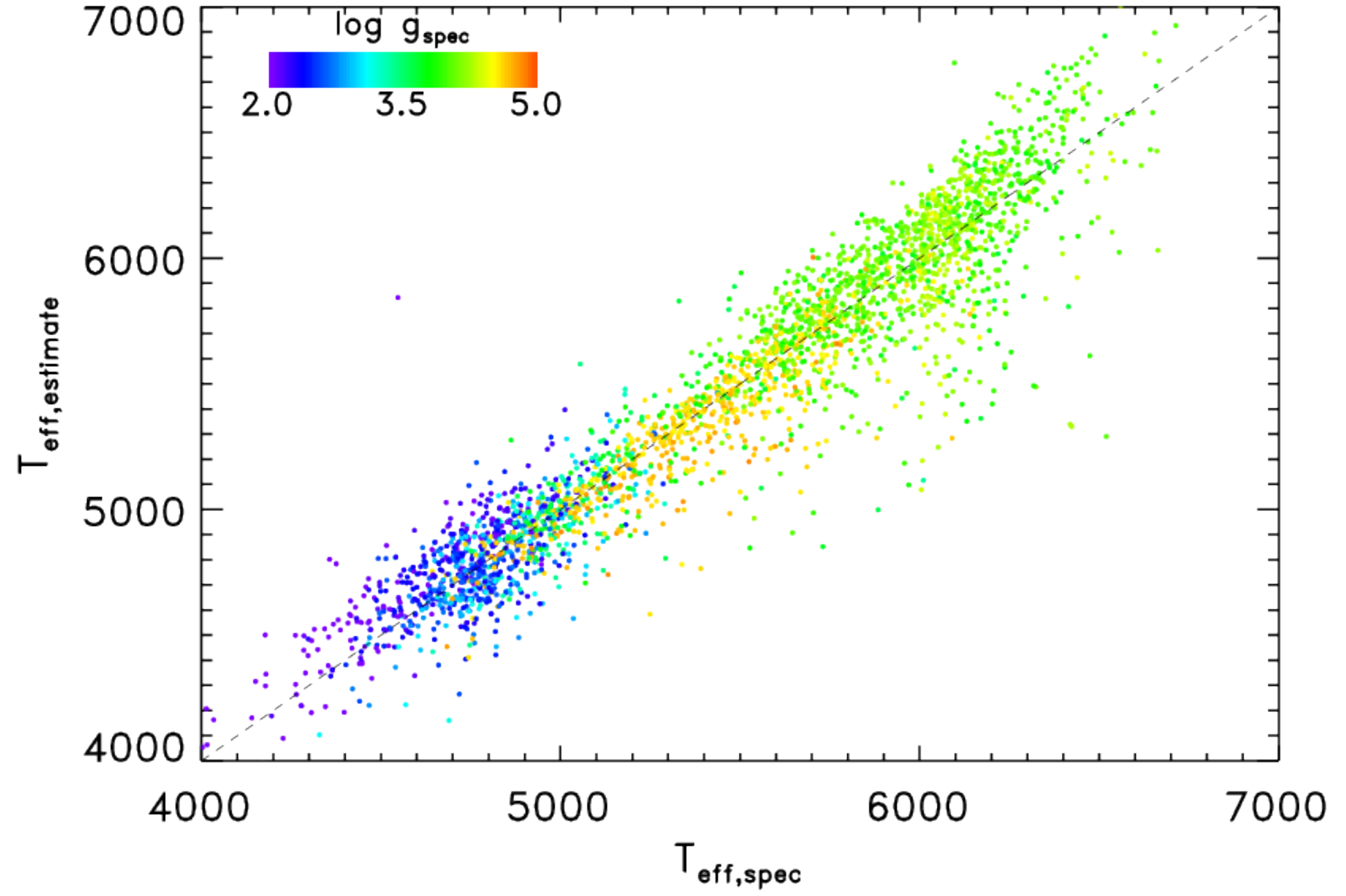}
&
\includegraphics[scale=0.27]{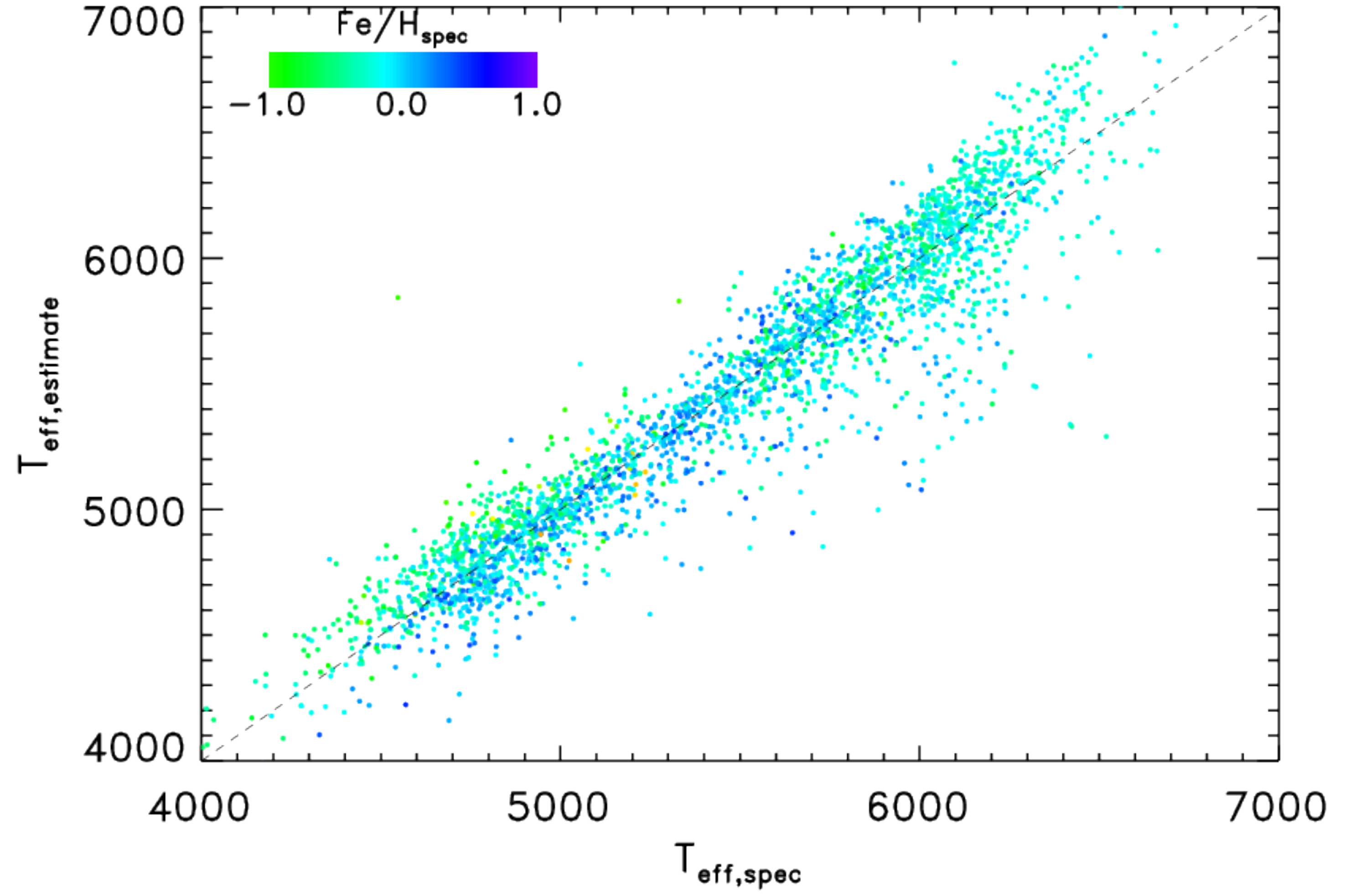}
\\
\includegraphics[scale=0.27]{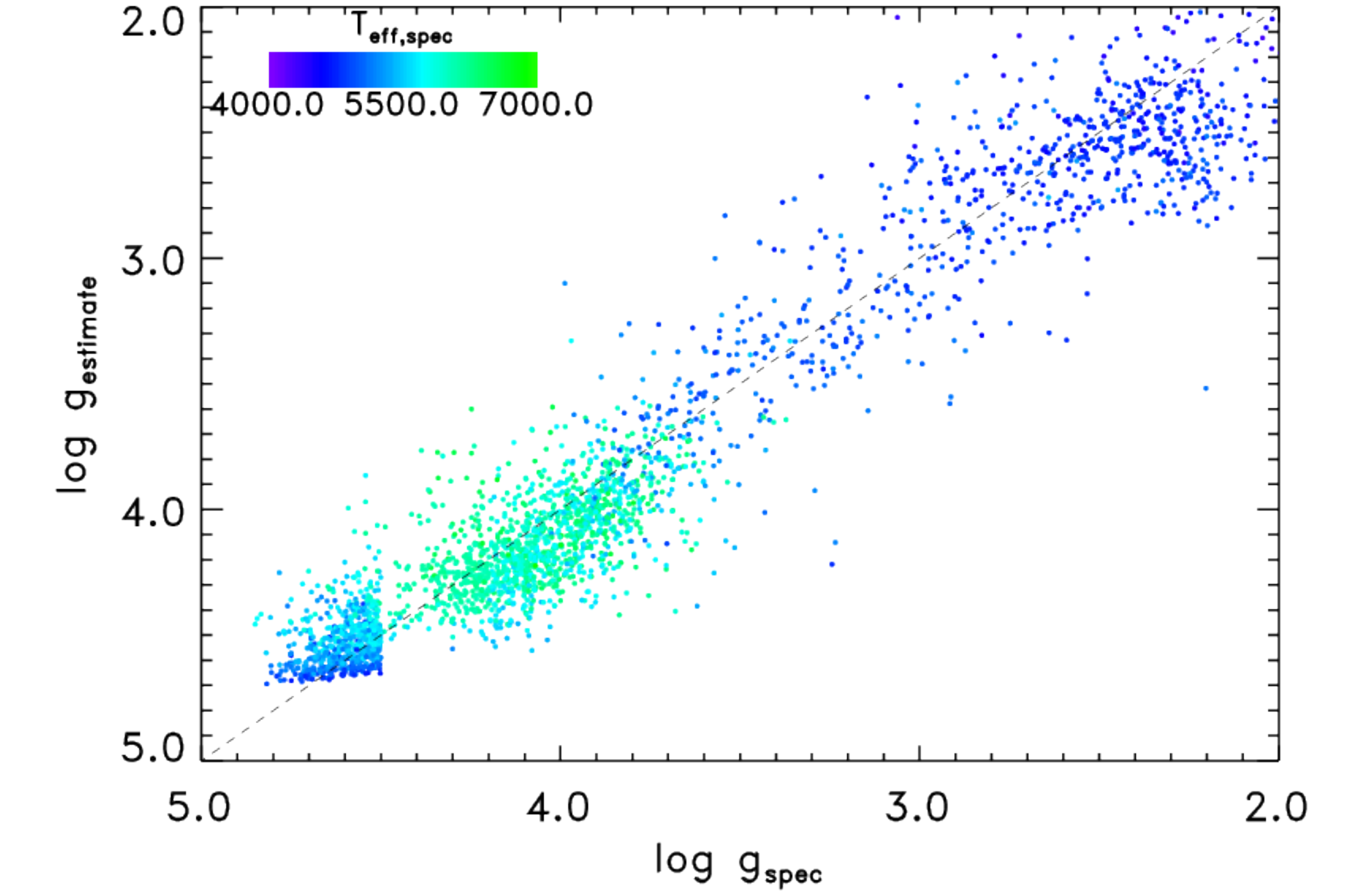}
&
\includegraphics[scale=0.27]{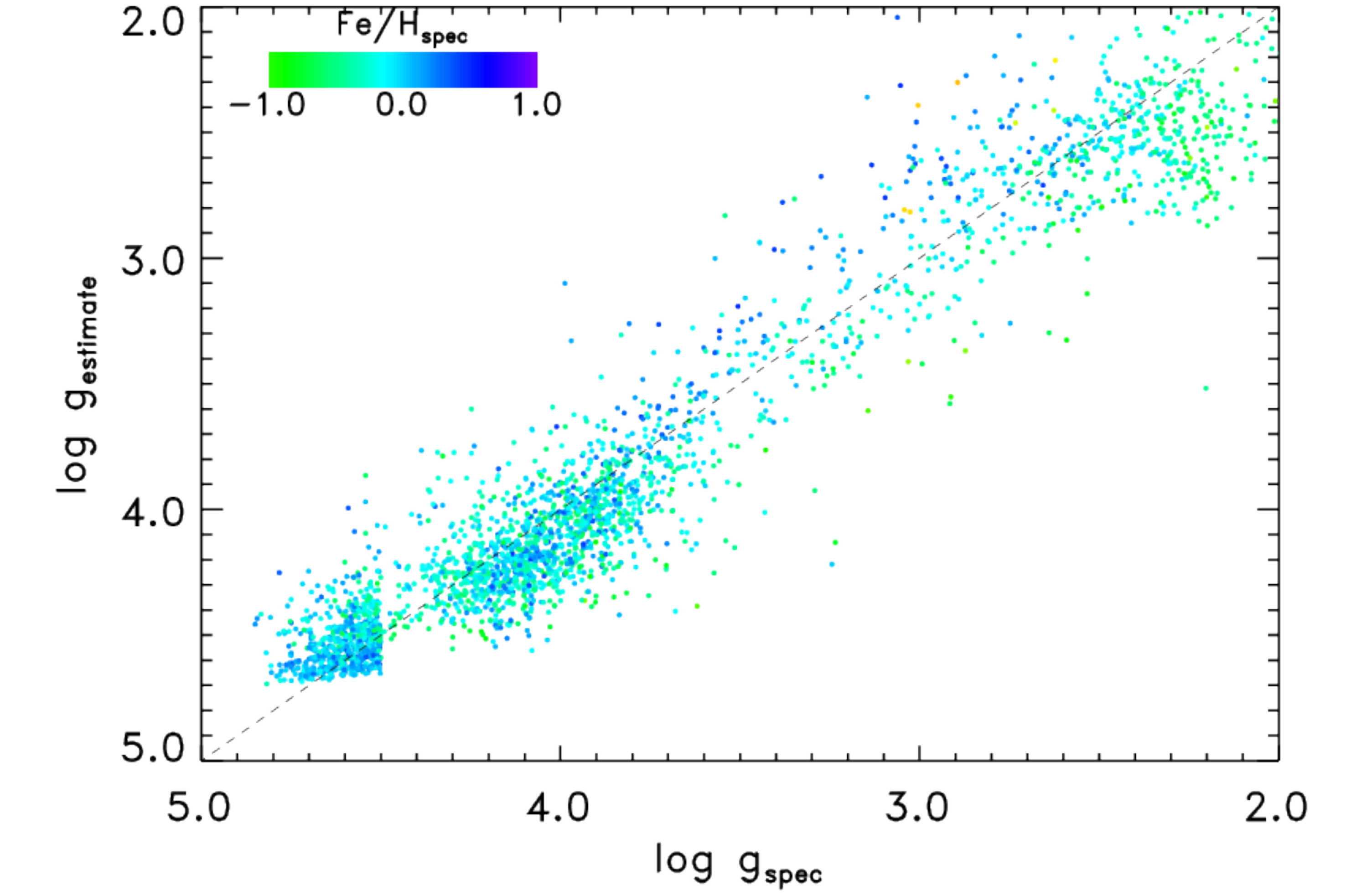}
\\
\includegraphics[scale=0.27]{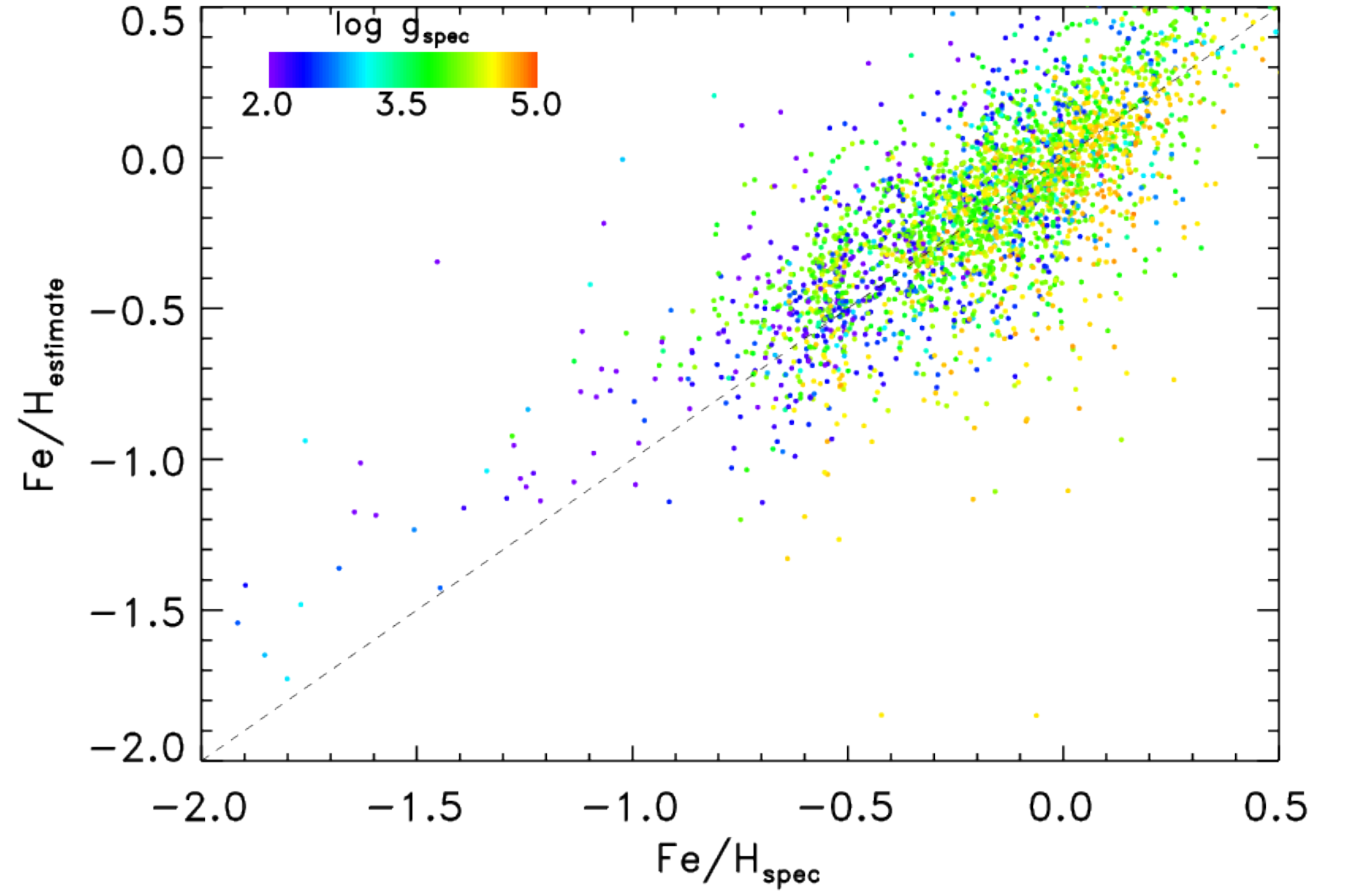}
&
\includegraphics[scale=0.27]{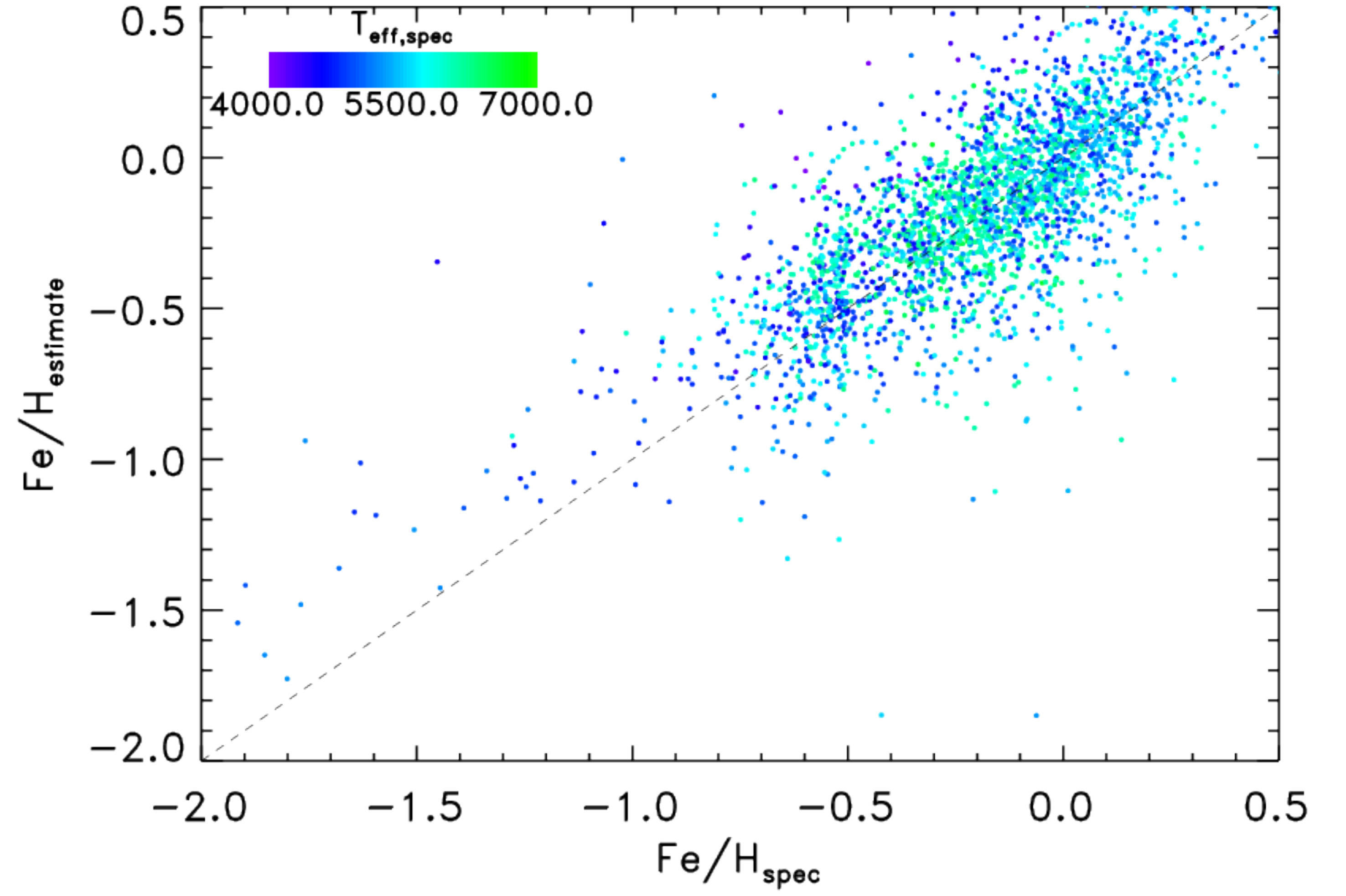}
\\

 \end{tabular}
 \caption[]{A comparison of the stellar parameters estimated from our method with those measured from GALAH spectroscopy for 2700 objects. We appear able to accurately reproduce effective temperature and gravity. For metallicity our method appears to work poorly for giants but relatively well for dwarfs. The structure at $\log g_{spec}$=4.5 is due to the selection of the sample used to make this plot, not the parameter estimation method. Note we truncate our colour scale at Fe/H=-1. The small number of objects more metal-poor than this will have a similar colour to an Fe/H=-1 object.} 
  \label{galah_params}
 \end{figure*}

We used a sample of 2700 GALAH stars to test how accurate our parameter estimation methods were. This sample was selected to cover a wide range of stellar parameters. We do not include objects hotter than 7000\,K and dwarfs colder than 4600\,K in our sample as our model was built on a training sample that included few such stars. The results of this parameter estimation test are shown in Figure~\ref{galah_params}.  We find that the mean value of $T_{eff,est}-T_{eff,spec}$ (the difference between our effective temperatures and the GALAH spectroscopic effective temperatures) is 9\,K. The scatter on the difference between these our effective temperature estimates and those quoted in the GALAH survey is 141\,K, broadly in line with what one would expect from GALAH effective temperatures with a typical uncertainty of 93\,K and our MCMC estimated effective temperatures with a typical uncertainty of 126\,K. This measurement uncertainty is larger than the offset between our temperature estimates and those in the TESS CTL. For metallicity we find that our estimate is comparable with the measured GALAH metallicities with an offset in $(Fe/H)_{eff,est}-(Fe/H)_{eff,spec}$ of 0.03 and a scatter of 0.20. This scatter is well below the typical uncertainty on our metallicity estimates of 0.29. Table~\ref{par_comp} contains a more detailed comparison of our stellar parameter estimates with those in the GALAH survey. There we break the results down into dwarfs of different spectral types and evolved stars. We tested our stellar parameter estimates using three combinations of the available photometry. First we used samples where all stars had good quality Skymapper data. We then set our code to treat either the blue Skymapper bands ($u$ and $v$) as missing or to treat all Skymapper bands as missing. This allowed us to test the accuracy of our method for the majority of stars where we had all the available photometry as well as the small number of stars with missing Skymapper data.

\begin{figure}
 \setlength{\unitlength}{1mm}
\includegraphics[scale=0.27]{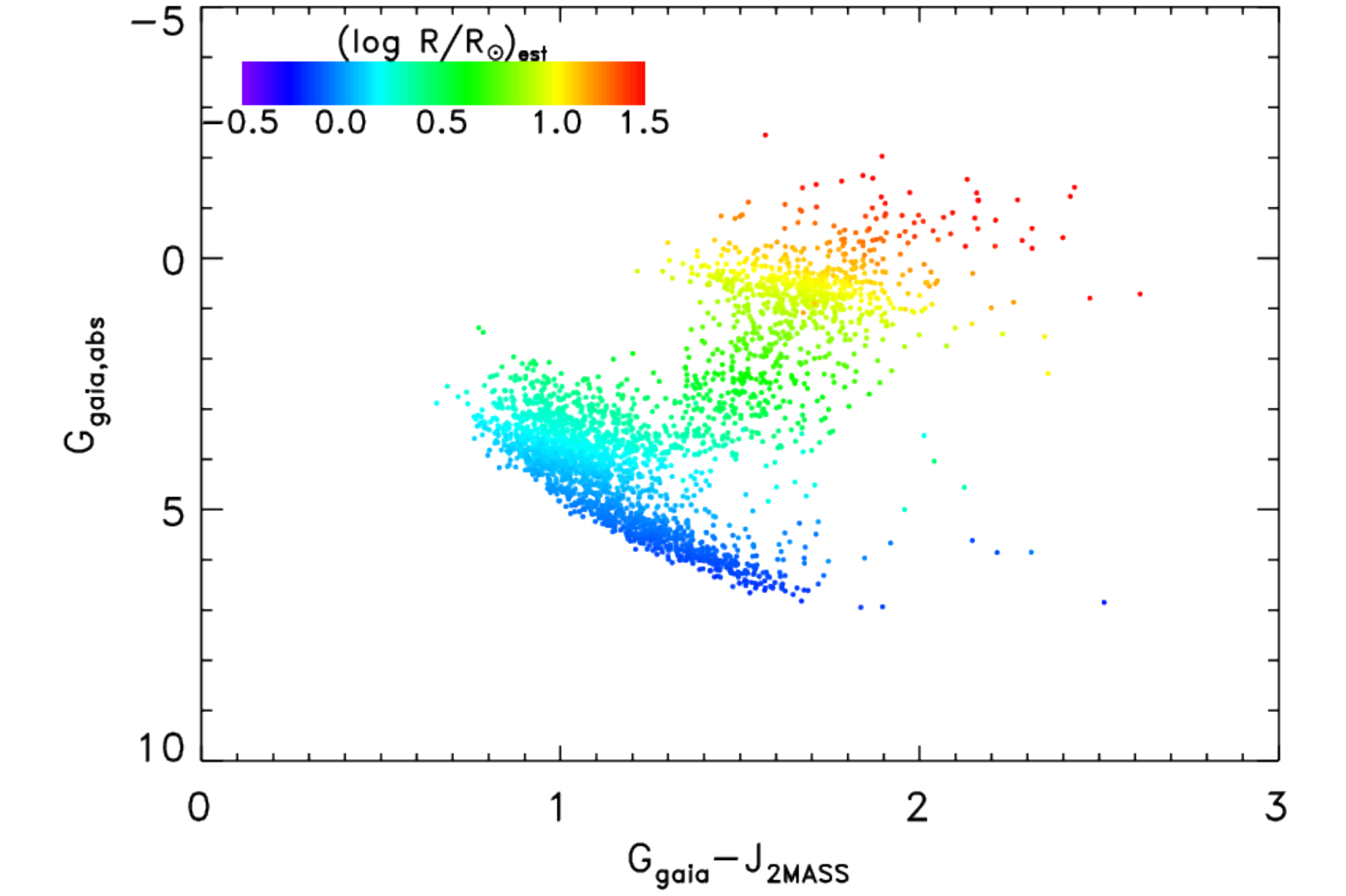}
 \caption[]{The estimated stellar radii for 2700 objects from the GALAH sample plotted on a colour-magnitude diagram. Note that the stellar radii increase up the subgiant branch as expected.} 
  \label{logr}
 \end{figure}

Figure~\ref{logr} shows the estimated stellar radii for our test sample. We successfully reproduce the expected pattern of larger, more luminous stars having larger radius estimates.
 
\subsubsection{Comparison to TESS Candidate Target List}
The TESS CTL (\citealt{Stassun2018}) contains stellar parameter estimates from a range of methods. These are a) spectroscopic measurements from large-scale surveys and b) where spectroscopic measurements are unavailable (i.e. most stars) effective temperatures from photometric colours. As Gaia~DR2 was unavailable when the CTL was originally constructed, not all stars have parallax measurements. \cite{Stassun2018} used reduced proper motion as a proxy for luminosity in order to separate dwarfs from giants. 

\begin{figure*}
 \setlength{\unitlength}{1mm}
 \begin{tabular}{cc}
\includegraphics[scale=0.27]{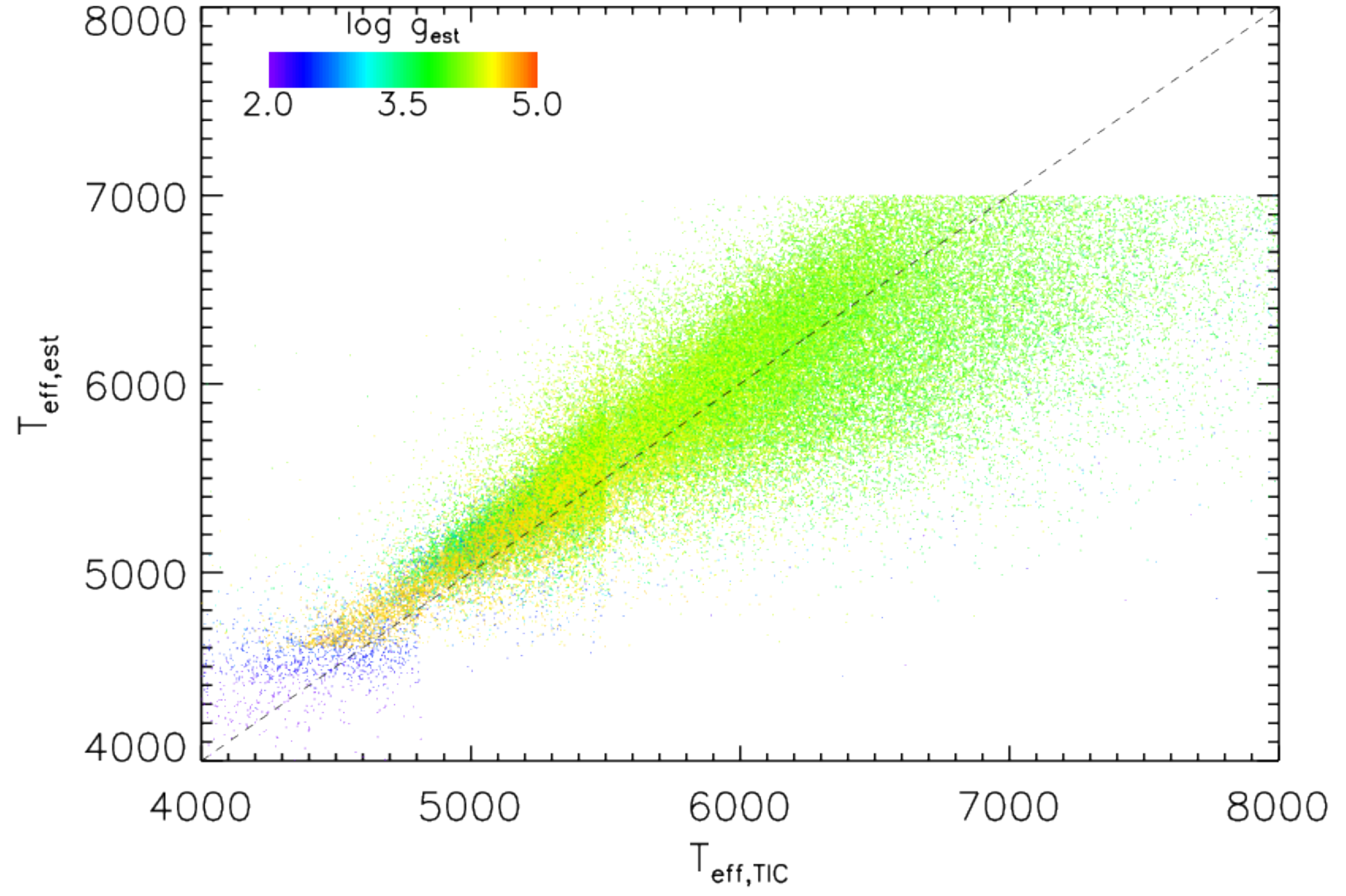}
&
\includegraphics[scale=0.27]{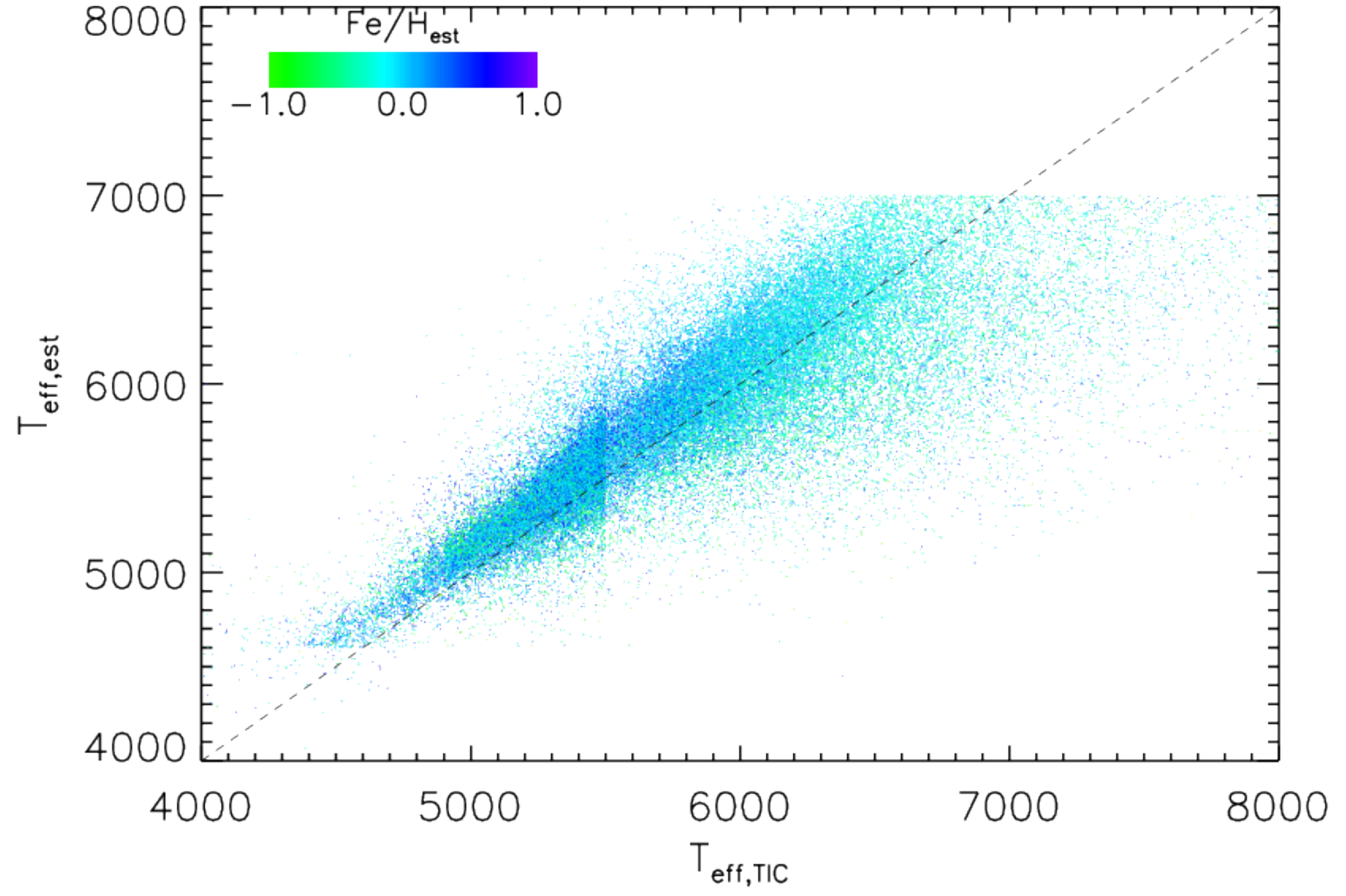}
\\ 
 \end{tabular}
 \caption[]{A comparison of the effective temperature estimates between our method and that of the TESS Input Catalog \protect\cite{Stassun2018}. The temperature estimates are broadly in agreement for stars from 4600\,K to 7000\,K.  Due to the large size of our sample, only 95,000 randomly selected stars are shown. Note the right-hand plot excludes stars which we could not determine the metallicity for as they had no blue Skymapper photometry. Note we truncate our colour scale at Fe/H=-1. The small number of objects more metal-poor than this will have a similar colour to an Fe/H=-1 object.} 
  \label{tic_comp}
 \end{figure*}

We compared our estimated effective temperatures to those in the CTL (see Figure~\ref{tic_comp}). We find that the temperatures are generally similar with our temperatures being on average 50\,K hotter. The scatter between the two measurements (calculated from 1.48 times the median absolute deviation) is 258\,K, slightly larger than the typical quoted effective temperature uncertainties in both our estimates and the CTL estimates 259\,K and 189\,K respectively). A large part of this scatter is due to objects in the Galactic Plane for which we have no blue Skymapper data and high limiting extinction values.  When we restrict ourselves to objects with a CTL priority above 0.001 (which excludes almost all stars in the Plane) we find a temperature bias of 111\,K and a scatter of 191\,K. Note both our method and the CTL use 2MASS data so are not entirely independent. The full comparison with the TESS CTL is shown in Table~\ref{par_comp}.

\begin{figure*}
 \setlength{\unitlength}{1mm}
 \begin{tabular}{cc}
\includegraphics[scale=0.27]{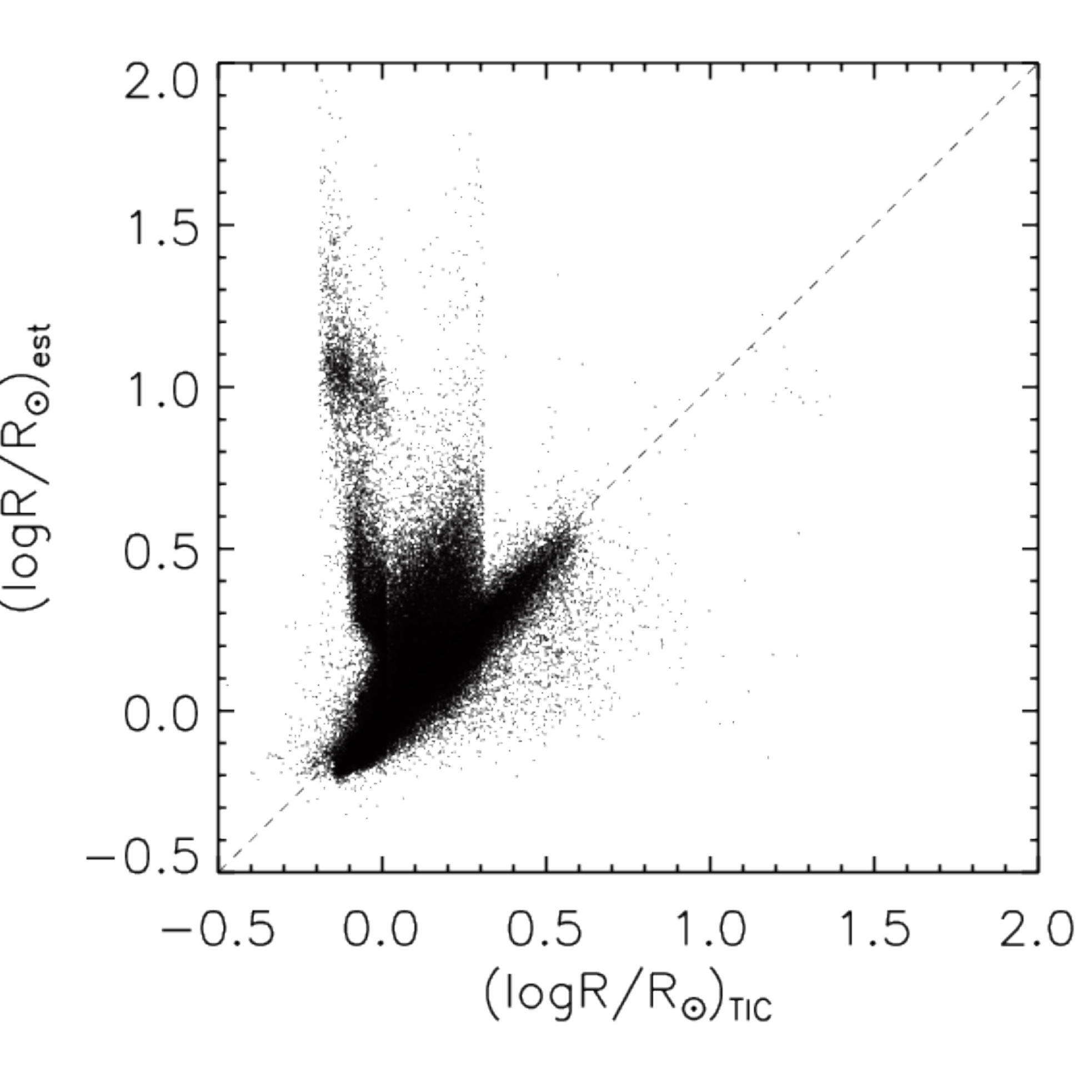}
&
\includegraphics[scale=0.27]{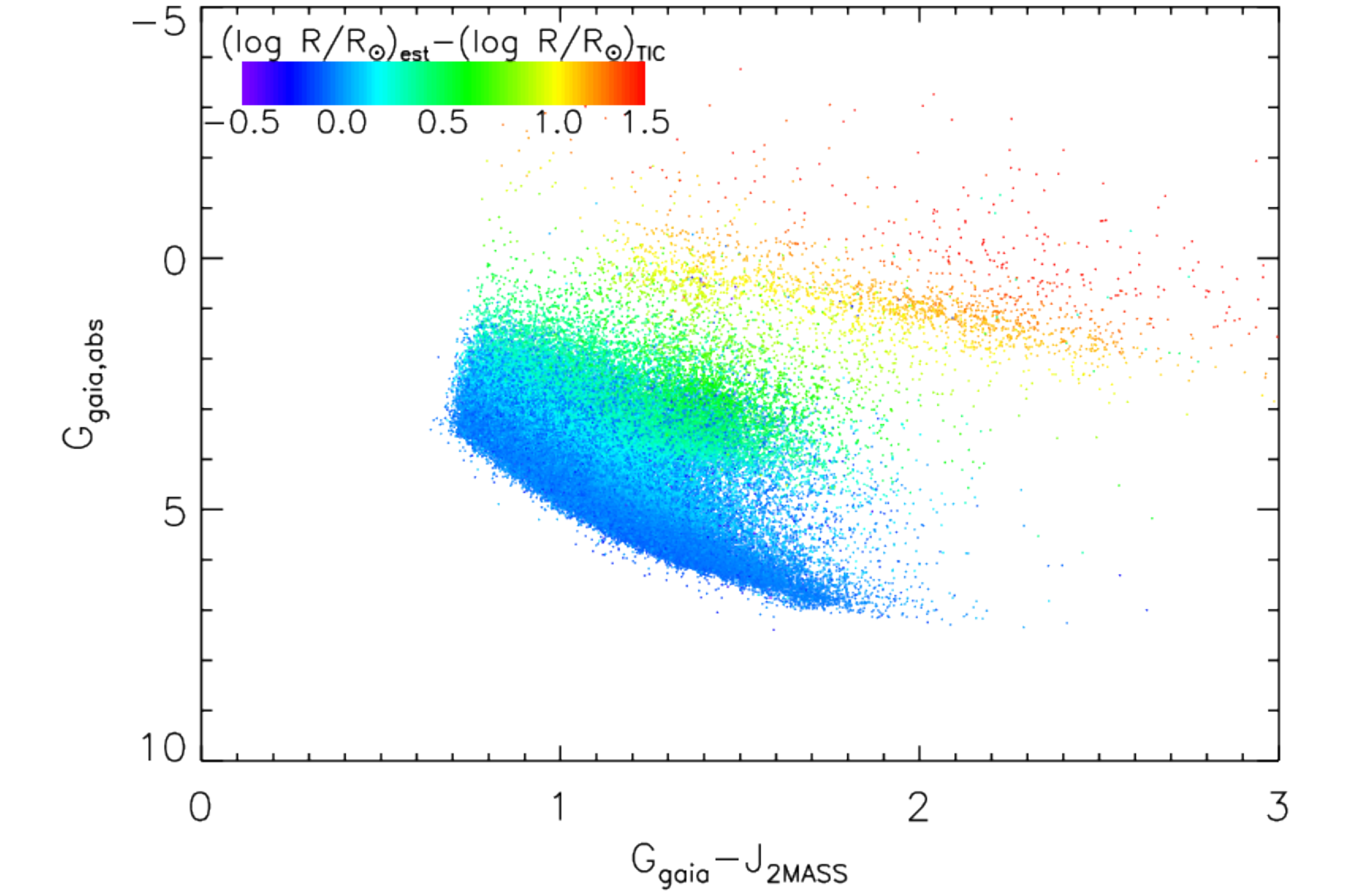}
\\ 
 \end{tabular}
 \caption[]{Left: A comparison of the stellar radius estimates between our method and that of the TESS CTL \protect\citep{Stassun2018}. There is one key point of disagreement, we estimate much larger radii for a small number of objects identified as dwarfs in the CTL. Right: The objects where we estimate larger radii are shown to be subgiants, not dwarfs. This is unsurprising as Gaia DR2 data was not available when the CTL was constructed. } 
  \label{logr_comp}
 \end{figure*}

We find that our stellar radius estimates are generally similar to the estimates in the CTL (see Figure~\ref{logr_comp}). There are relatively few giants as many of these were excluded by the cuts made by \cite{Stassun2018}. However we find that 22\% of our stars have stellar radius estimates that are 50\% larger than the estimates in the CTL. This reduces to 14\% for higher priority TESS targets (CTL priority$>$0.001). This is likely due to us having Gaia DR2 parallaxes available to us, allowing us to better estimate the absolute magnitudes of stars. We note that the recently released CTL version~0702 uses Gaia DR2 parallaxes to remove some giant stars. This changes our previously mentioned contamination numbers to 21\% for the full catalogue and 14\% for high priority targets. The right-hand panel of Figure~\ref{logr_comp} shows that these stars with larger radius estimates in our study are typically subgiants. Note that the radius uncertainties in the CTL takes into account the fact that it is difficult to distinguish a dwarf from a subgiant using reduced proper motion. The radius uncertainty is capped at 100\% of the radius estimate. As a result, the CTL median fractional radius uncertainty (the final column of Table~\ref{par_comp}) can either be unity or close to unity. 

\begin{figure}
 \setlength{\unitlength}{1mm}

\includegraphics[scale=0.27]{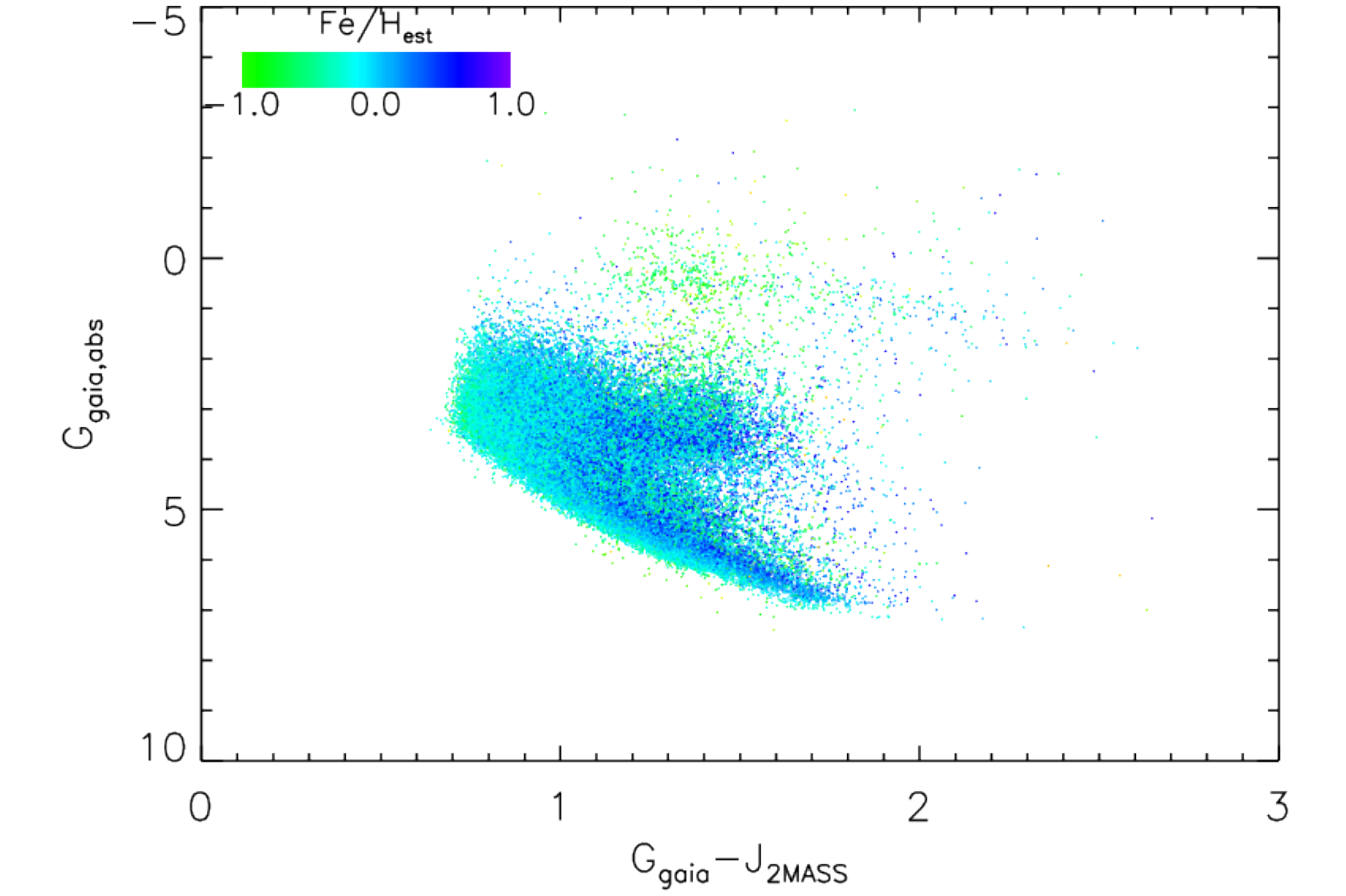}
 \caption[]{A colour-magnitiude diagram showing the metallicity estimates from our method. Broadly we see the expected pattern of metal-poor stars appearing underluminous. We see an excess of metal-poor stars among subgiants and above the main sequence. This may be the result of our method not considering binarity. Note this plot excludes stars which we could not determine the metallicity for as they had no blue Skymapper photometry. Note we truncate our colour scale at Fe/H=-1. The small number of objects more metal-poor than this will have a similar colour to an Fe/H=-1 object.}
  \label{feh_cmd}
 \end{figure}
\subsubsection{Comparison to other stellar parameter estimates}
\label{comp_sect}
To further test our stellar parameter measurements we also compared our results with objects in the GALAH-TGAS survey (a separate stellar parameter calculation from the main GALAH survey using GALAH spectra and GAIA TGAS parallaxes \protect\citealt{Buder2018a}), K2 asteroseismology parameters \protect\citep{Lund2016} and the aforementioned TESS Input Catalog \protect\citep{Stassun2018}. We note that the GALAH-TGAS survey includes some of the astrometric data that went into the Gaia parallaxes we use, hence it is not entirely independent. Our comparison with the K2 asteroseismology stars was done without Skymapper data as the stars in this comparison sample are almost all too bright to have reliable Skymapper photometry. Note that as we do not estimate metallicity for stars with no blue Skymapper photometry, comparison runs where we remove the either the blue Skymapper photometry or all the Skymapper photometry have no metallicity measurement comparison.

Table~\ref{par_comp} shows all our parameter comparisons, including our comparisons with GALAH and the TESS CTL. It appears that our method has small biases for all parameters compared to our comparison samples. Our uncertainties typically agree well with the measured scatter between our measurements and those from other surveys. The typical fractional radius estimate in our comparison with asteroseismology stars is a few percent higher.
\subsection{Limitations to our model}
To test the potential limitations of our model we examine Figures~\ref{logr_comp} and~\ref{feh_cmd}. We note a small number of objects that lie between the main sequence and giant branch which have very high metallicity estimates (typically over 0.55\,dex). Such objects are either overluminous binaries or objects with red blue-optical colours which our model cannot fit well. We flag these objects in our catalogue.

We find that for the the vast majority of our objects we do not constrain extinction. The likelihood for each of these objects result is roughly flat between our limiting extinction values. This is typically because most of our objects have limiting extinctions of 0.1-0.2\,mag. The reddening in each of our colours will be less than this (and highest in $G-W_2$) while the scatter we measure on each colour is typically 0.1\,mag. This means that for most stars we cannot constrain the extinction well. Hence for objects with estimated extinction values less significant than 5$\sigma$ we simply quote our limiting extinction as an upper limit.

\subsection{Our catalogue}
A short example table of our catalogue is shown in Table~\ref{short_cat}. We quote values for effective temperature, gravity, metallicity, radius, mass and extinction along with one standard deviation error bounds. We do not quote an age estimate as this is relatively unconstrained for stars on the main sequence. As stated above, we only quote extinction values for stars where our extinction values are more than five times the quoted extinction uncertainty. For objects where one of our error bounds fall outside the range of our model we quote lower limits on the uncertainty. For objects where we did not estimate the metallicity we quote $-$0.1\,dex as a typical value in the solar neighbourhood and flag these values with a {$\dag$}.

\section{Conclusions}
We have calculated stellar parameters for 939,457 FGK stars that are likely to be short-cadence targets for the TESS mission. Our method uses data-driven models for colours which accurately reproduce the observed colours of sun-like stars. We are able to estimate the radii of TESS targets with a typical uncertainty of 9.3\%. This catalogue can be used to screen exoplanet candidates from TESS and provides a homogeneous set of stellar parameters for statistical studies.
\section*{Acknowledgments}
This work presents results from the European Space Agency (ESA) space mission Gaia. Gaia data are being processed by the Gaia Data Processing and Analysis Consortium (DPAC). Funding for the DPAC is provided by national institutions, in particular the institutions participating in the Gaia MultiLateral Agreement (MLA). The Gaia mission website is https://www.cosmos.esa.int/gaia. The Gaia archive website is https://archives.esac.esa.int/gaia. This publication makes use of data products from the Wide-field Infrared Survey Explorer, which is a joint project of the University of California, Los Angeles, and the Jet Propulsion Laboratory/California Institute of Technology, and NEOWISE, which is a project of the Jet Propulsion Laboratory/California Institute of Technology. WISE and NEOWISE are funded by the National Aeronautics and Space Administration. The national facility capability for SkyMapper has been funded through ARC LIEF grant LE130100104 from the Australian Research Council, awarded to the University of Sydney, the Australian National University, Swinburne University of Technology, the University of Queensland, the University of Western Australia, the University of Melbourne, Curtin University of Technology, Monash University and the Australian Astronomical Observatory. SkyMapper is owned and operated by The Australian National University's Research School of Astronomy and Astrophysics. The survey data were processed and provided by the SkyMapper Team at ANU. The SkyMapper node of the All-Sky Virtual Observatory (ASVO) is hosted at the National Computational Infrastructure (NCI). Development and support the SkyMapper node of the ASVO has been funded in part by Astronomy Australia Limited (AAL) and the Australian Government through the Commonwealth's Education Investment Fund (EIF) and National Collaborative Research Infrastructure Strategy (NCRIS), particularly the National eResearch Collaboration Tools and Resources (NeCTAR) and the Australian National Data Service Projects (ANDS). N.R.D acknowledges acknowledges the support of the DFG priority program SPP 1992 "Exploring the Diversity of Extrasolar Planets (Characterising the population of wide orbit exoplanets)". The authors thank Morgan Fouesneau for helpful discussions and Wolfgang Brandner for manuscript comments. 

\bibliographystyle{mn2e} 
 \bibliography{./ndeacon} 

\begin{landscape}
\begin{table*}[h]
\caption{Comparisons of our measurements with those from other works: the GALAH survey \protect\citep{Buder2018}, the GALAH-TGAS survey \protect\citep{Buder2018a}, K2 asteroseismology parameters \protect\citep{Lund2016} and the TESS Input Catalog \protect\citep{Stassun2018}. We for all but the asteroseismology comparison we divide our sample into evolved stars (where the comparison survey measures $\log g<4$) and dwarfs of different spectral classes. We follow \protect\cite{Deacon2016} and use the effective temperature scale of \protect\cite{Pecaut2013} to define the boundaries of each spectral class. For each parameter $\Delta x$ is the median difference of the survey measurements, $\sigma_{x}$ is the robust scatter between the survey measurements (i.e. 1.48 times the median absolute deviation), $\sigma_{x,m}$ is the mean measurement uncertainty in our work and $\sigma_{x,c}$ is the mean measurement uncertainty in the comparison work. The "Data Used" column details which data were used in which calculation. We tested our method with all stars having Skymapper data ("All"), ignoring blue Skymapper data ("NoBlue") and ignoring all Skymapper data ("NoOpt"). For the TESS CTL comparison we used whatever photometry was available for each object.}
\label{par_comp}

\tiny
\begin{center}
\begin{tabular}{rrrrrrrrrrrrrrrrrrrrrr}
\hline
&Data&$\Delta_{T}$&$\sigma_{T}$&$\sigma_{T,m}$&$\sigma_{T,c}$&$\Delta_{\log g}$&$\sigma_{\log g}$&$\sigma_{\log g,m}$&$\sigma_{\log g,c}$&$\Delta_{Fe/H}$&$\sigma_{Fe/H}$&$\sigma_{Fe/H,m}$&$\sigma_{Fe/H,c}$&$\frac{\Delta_{M}}{M}$&$\frac{\sigma_{M}}{M}$&$\frac{\sigma_{M,m}}{M}$&$\frac{\sigma_{M,c}}{M}$&$\frac{\Delta_{R}}{R}$&$\frac{\sigma_{R}}{R}$&$\frac{\sigma_{R,m}}{R}$&$\frac{\sigma_{R,c}}{R}$\\
&used&\multicolumn{4}{c}{(K)}&\multicolumn{4}{c}{(dex.)}&\multicolumn{4}{c}{(dex.)}\\
\hline
\multicolumn{14}{c}{GALAH survey}\\
\hline
F Dwarf&All&20&184&198& 69&0.01&0.15&0.12&0.19&0.02&0.18&0.27&0.08&\ldots&\ldots&\ldots&\ldots&\ldots&\ldots&\ldots&\ldots\\
G Dwarf&All&1&136&140& 74&$-$0.02&0.16&0.09&0.19&$-$0.01&0.20&0.30&0.08&\ldots&\ldots&\ldots&\ldots&\ldots&\ldots&\ldots&\ldots\\
K Dwarf&All&$-$14& 74& 67& 70&$-$0.02&0.10&0.06&0.19&$-$0.02&0.14&0.23&0.08&\ldots&\ldots&\ldots&\ldots&\ldots&\ldots&\ldots&\ldots\\
Evolved&All&22&145&107& 72&0.09&0.21&0.16&0.19&0.06&0.21&0.31&0.08&\ldots&\ldots&\ldots&\ldots&\ldots&\ldots&\ldots&\ldots\\
F Dwarf&NoBlue&40&203&188& 69&0.04&0.14&0.09&0.19&\ldots&\ldots&\ldots&\ldots&\ldots&\ldots&\ldots&\ldots&\ldots&\ldots&\ldots&\ldots\\
G Dwarf&NoBlue&10&145&134& 74&$-$0.03&0.12&0.08&0.19&\ldots&\ldots&\ldots&\ldots&\ldots&\ldots&\ldots&\ldots&\ldots&\ldots&\ldots&\ldots\\
K Dwarf&NoBlue&1& 89& 91& 70&$-$0.03&0.09&0.05&0.19&\ldots&\ldots&\ldots&\ldots&\ldots&\ldots&\ldots&\ldots&\ldots&\ldots&\ldots&\ldots\\
Evolved&NoBlue&9&160&113& 72&0.09&0.23&0.12&0.19&\ldots&\ldots&\ldots&\ldots&\ldots&\ldots&\ldots&\ldots&\ldots&\ldots&\ldots&\ldots\\
F Dwarf&NoOpt.&34&188&200& 69&0.03&0.13&0.10&0.19&\ldots&\ldots&\ldots&\ldots&\ldots&\ldots&\ldots&\ldots&\ldots&\ldots&\ldots&\ldots\\
G Dwarf&NoOpt.&21&138&144& 74&$-$0.03&0.13&0.08&0.19&\ldots&\ldots&\ldots&\ldots&\ldots&\ldots&\ldots&\ldots&\ldots&\ldots&\ldots&\ldots\\
K Dwarfs&NoOpt.&4& 86& 98& 70&$-$0.03&0.09&0.05&0.19&\ldots&\ldots&\ldots&\ldots&\ldots&\ldots&\ldots&\ldots&\ldots&\ldots&\ldots&\ldots\\
Evolved&NoOpt.&14&153&122& 72&0.11&0.23&0.13&0.19&\ldots&\ldots&\ldots&\ldots&\ldots&\ldots&\ldots&\ldots&\ldots&\ldots&\ldots&\ldots\\
\hline
\multicolumn{14}{c}{GALAH$-$TGAS survey}\\
\hline
F Dwarf&All&98&150&309& 94&0.08&0.10&0.12&0.20&$-$0.06&0.17&0.26&0.07&0.00&0.05&0.17&0.05&\ldots&\ldots&\ldots&\ldots\\
G Dwarf&All&57&113&203& 94&0.09&0.09&0.11&0.15&$-$0.02&0.16&0.29&0.07&0.02&0.05&0.16&0.05&\ldots&\ldots&\ldots&\ldots\\
K Dwarf&All&46&106& 56& 94&0.05&0.04&0.05&0.10&$-$0.04&0.15&0.19&0.08&0.02&0.04&0.10&0.04&\ldots&\ldots&\ldots&\ldots\\
F Dwarf&NoBlue&111&169&317& 94&0.11&0.11&0.10&0.20&\ldots&\ldots&\ldots&\ldots&0.02&0.04&0.11&0.05&\ldots&\ldots&\ldots&\ldots\\
G Dwarf&NoBlue&68&117&217& 94&0.07&0.10&0.10&0.15&\ldots&\ldots&&\ldots\ldots&0.01&0.05&0.11&0.05&\ldots&\ldots&\ldots&\ldots\\
K Dwarf&No Blue&63&106&104& 94&0.05&0.03&0.04&0.10&\ldots&\ldots&\ldots&\ldots&0.01&0.03&0.05&0.0&\ldots&\ldots&\ldots&\ldots\\
F Dwarf&NoOpt.&117&170&332& 94&0.11&0.10&0.11&0.20&\ldots&\ldots&\ldots&\ldots&0.02&0.04&0.11&0.05&\ldots&\ldots&\ldots&\ldots\\
G Dwarf&NoOpt.&71&118&229& 94&0.08&0.10&0.10&0.15&\ldots&\ldots&\ldots&\ldots&0.01&0.05&0.11&0.05&\ldots&\ldots&\ldots&\ldots\\
K Dwarfs&NoOpt.&55&109&125& 94&0.05&0.04&0.05&0.10&\ldots&\ldots&\ldots&\ldots&0.01&0.03&0.05&0.04&\ldots&\ldots&\ldots&\ldots\\

\hline
\multicolumn{14}{c}{K2 Asteroseismology stars}\\
\hline
F/G Dwarfs&NoOpt.&\ldots&\ldots&\ldots&\ldots&0.07&0.09&0.12&0.01&\ldots&\ldots&\ldots&\ldots&0.07&0.07&0.11&0.04&0.00&0.10&0.13&0.02\\
\hline
\multicolumn{14}{c}{TESS CTL}\\
\hline
F Dwarf&&84&313&318&195&\ldots&\ldots&\ldots&\ldots&\ldots&\ldots&\ldots&\ldots&\ldots&\ldots&\ldots&\ldots&$-$0.03&0.06&0.10&0.31\\
G Dwarf&&64&207&188&186&\ldots&\ldots&\ldots&\ldots&\ldots&\ldots&\ldots&\ldots&\ldots&\ldots&\ldots&\ldots&$-$0.03&0.07&0.08&1.00\\
K Dwarf&&58&143&106&181&\ldots&\ldots&\ldots&\ldots&\ldots&\ldots&\ldots&\ldots&\ldots&\ldots&\ldots&\ldots&$-$0.05&0.04&0.07&1.00\\
Evolved&&$-$20&347&371&191&\ldots&\ldots&\ldots&\ldots&\ldots&\ldots&\ldots&\ldots&\ldots&\ldots&\ldots&\ldots&0.22&0.24&0.11&0.97\\

F Dwarf High&&177&212&246&192&\ldots&\ldots&\ldots&\ldots&\ldots&\ldots&\ldots&\ldots&\ldots&\ldots&\ldots&\ldots&$-$0.03&0.04&0.09&0.16\\
G Dwarf High&&75&176&142&186&\ldots&\ldots&\ldots&\ldots&\ldots&\ldots&\ldots&\ldots&\ldots&\ldots&\ldots&\ldots&$-$0.03&0.04&0.08&0.18\\
K Dwarf High&&88&121& 72&180&\ldots&\ldots&\ldots&\ldots&\ldots&\ldots&\ldots&\ldots&\ldots&\ldots&\ldots&\ldots&$-$0.05&0.04&0.06&0.20\\
Evolved High&&109&202&282&186&\ldots&\ldots&\ldots&\ldots&\ldots&\ldots&\ldots&\ldots&\ldots&\ldots&\ldots&\ldots&0.26&0.25&0.10&1.00\\
\hline
\end{tabular}
\end{center}
\normalsize
\end{table*}%
\end{landscape}
\begin{table*}[h]
\caption{A short example table of our stellar parameter estimates for TESS targets.}
\begin{center}
\begin{tabular}{rlrrrrr}
\hline
TESS ID&$T_{eff,est}$(K)&$\log g_{est}$(dex)&$(Fe/H)_{est}$(dex)&$\log_{10}M_{est}/M_{\odot}$&$\log_{10} R_{est}/R_{\odot}$&$A_{V,est}$(mag)\\
\hline
101&6099$\pm^{ 279 }_{ 117 }$&4.42$\pm^{ 0.07 }_{ 0.13 }$&-0.07$\pm^{ 0.20 }_{ 0.28 }$&0.02$\pm^{ 0.06 }_{ 0.08 }$&0.02$\pm^{ 0.04 }_{ 0.03 }$&$<$0.251\\ 
159&6010$\pm^{ 339 }_{ 119 }$&4.37$\pm^{ 0.10 }_{ 0.13 }$&0.05$\pm^{ 0.25 }_{ 0.29 }$&0.03$\pm^{ 0.07 }_{ 0.07 }$&0.05$\pm^{ 0.04 }_{ 0.04 }$&$<$0.261\\ 
164&5503$\pm^{ 266 }_{ 147 }$&4.44$\pm^{ 0.14 }_{ 0.05 }$&-0.10$\pm^{ 0.34 }_{ 0.43 }$&-0.08$\pm^{ 0.11 }_{ 0.02 }$&-0.04$\pm^{ 0.04 }_{ 0.04 }$&$<$0.230\\ 
185&5649$\pm^{ 253 }_{ 86 }$&4.41$\pm^{ 0.11 }_{ 0.10 }$&0.15$\pm^{ 0.30 }_{ 0.23 }$&-0.01$\pm^{ 0.08 }_{ 0.06 }$&0.01$\pm^{ 0.03 }_{ 0.04 }$&$<$0.281\\ 
358&$>$7000&\ldots&\ldots&\ldots&\ldots&\ldots\\ 
382&6159$\pm^{ 676 }_{ 101 }$&4.08$\pm^{ 0.20 }_{ 0.09 }$&0.04$\pm^{ 0.31 }_{ 0.25 }$&0.08$\pm^{ 0.13 }_{ 0.05 }$&0.22$\pm^{ 0.04 }_{ 0.05 }$&$<$0.278\\ 
479&5458$\pm^{ 220 }_{ 74 }$&4.12$\pm^{ 0.04 }_{ 0.05 }$&0.49$\pm^{ 0.16 }_{ 0.26 }$&0.20$\pm^{ 0.02 }_{ 0.05 }$&0.26$\pm^{ 0.04 }_{ 0.04 }$&$<$0.270\\ 
1178&5417$\pm^{ 483 }_{ 123 }$&4.42$\pm^{ 0.10 }_{ 0.09 }$&-0.10$\pm^{ 0.00 }_{ 0.00 }$&-0.02$\pm^{ 0.03 }_{ 0.06 }$&0.00$\pm^{ 0.03 }_{ 0.05 }$&$<$0.303\\ 
1238&5999$\pm^{ 404 }_{ 191 }$&4.32$\pm^{ 0.11 }_{ 0.16 }$&0.12$\pm^{ 0.36 }_{ 0.42 }$&0.04$\pm^{ 0.09 }_{ 0.10 }$&0.08$\pm^{ 0.04 }_{ 0.04 }$&$<$0.309\\ 
1275&5932$\pm^{ 200 }_{ 157 }$&4.51$\pm^{ 0.03 }_{ 0.18 }$&-0.01$\pm^{ 0.17 }_{ 0.40 }$&0.02$\pm^{ 0.03 }_{ 0.12 }$&-0.02$\pm^{ 0.04 }_{ 0.03 }$&$<$0.149\\ 
1307&5341$\pm^{ 462 }_{ 74 }$&4.50$\pm^{ 0.10 }_{ 0.07 }$&-0.10$\pm^{ 0.00 }_{ 0.00 }$&-0.06$\pm^{ 0.06 }_{ 0.04 }$&-0.06$\pm^{ 0.02 }_{ 0.04 }$&$<$0.314\\ 
1325&6896$\pm^{ >103 }_{ 229 }$&4.18$\pm^{ 0.20 }_{ 0.13 }$&-0.23$\pm^{ 0.33 }_{ 0.46 }$&0.12$\pm^{ 0.08 }_{ 0.07 }$&0.19$\pm^{ 0.04 }_{ 0.08 }$&$<$0.317\\ 
1420&6425$\pm^{ >574 }_{ 146 }$&3.85$\pm^{ 0.24 }_{ 0.12 }$&-0.16$\pm^{ 0.41 }_{ 0.32 }$&0.16$\pm^{ 0.08 }_{ 0.06 }$&0.38$\pm^{ 0.04 }_{ 0.09 }$&$<$0.324\\ 
1586&5470$\pm^{ 239 }_{ 102 }$&4.50$\pm^{ 0.11 }_{ 0.06 }$&-0.17$\pm^{ 0.34 }_{ 0.36 }$&-0.09$\pm^{ 0.09 }_{ 0.03 }$&-0.08$\pm^{ 0.03 }_{ 0.03 }$&$<$0.295\\ 
1648&6108$\pm^{ 604 }_{ 154 }$&4.28$\pm^{ 0.15 }_{ 0.09 }$&-0.10$\pm^{ 0.00 }_{ 0.00 }$&0.02$\pm^{ 0.09 }_{ 0.02 }$&0.09$\pm^{ 0.04 }_{ 0.05 }$&$<$0.312\\ 
1886&6790$\pm^{ >209 }_{ 182 }$&4.09$\pm^{ 0.22 }_{ 0.12 }$&-0.07$\pm^{ 0.35 }_{ 0.46 }$&0.15$\pm^{ 0.09 }_{ 0.07 }$&0.25$\pm^{ 0.04 }_{ 0.09 }$&$<$0.276\\ 
1919&5511$\pm^{ 149 }_{ 139 }$&4.54$\pm^{ 0.03 }_{ 0.15 }$&0.24$\pm^{ 0.20 }_{ 0.46 }$&0.01$\pm^{ 0.02 }_{ 0.10 }$&-0.05$\pm^{ 0.04 }_{ 0.03 }$&$<$0.278\\ 
1981&5695$\pm^{ 460 }_{ 94 }$&4.06$\pm^{ 0.20 }_{ 0.09 }$&0.38$\pm^{ 0.27 }_{ 0.37 }$&0.07$\pm^{ 0.12 }_{ 0.05 }$&0.22$\pm^{ 0.04 }_{ 0.05 }$&$<$0.294\\ 
2082&$<$4600&\ldots&\ldots&\ldots&\ldots&\ldots\\ 
2216&5367$\pm^{ 146 }_{ 194 }$&4.56$\pm^{ 0.04 }_{ 0.12 }$&0.19$\pm^{ 0.21 }_{ 0.63 }$&-0.02$\pm^{ 0.04 }_{ 0.09 }$&-0.07$\pm^{ 0.04 }_{ 0.03 }$&$<$0.285\\ 
2328&5556$\pm^{ 383 }_{ 212 }$&4.41$\pm^{ 0.15 }_{ 0.05 }$&-0.10$\pm^{ 0.00 }_{ 0.00 }$&-0.08$\pm^{ 0.09 }_{ 0.01 }$&-0.02$\pm^{ 0.04 }_{ 0.04 }$&$<$0.291\\ 
2503&5173$\pm^{ 111 }_{ 61 }$&4.62$\pm^{ 0.03 }_{ 0.10 }$&0.07$\pm^{ 0.14 }_{ 0.41 }$&-0.06$\pm^{ 0.02 }_{ 0.08 }$&-0.12$\pm^{ 0.03 }_{ 0.03 }$&$<$0.180\\ 
2700&5826$\pm^{ 640 }_{ 117 }$&4.12$\pm^{ 0.16 }_{ 0.13 }$&0.34$\pm^{ 0.31 }_{ 0.38 }$&0.10$\pm^{ 0.10 }_{ 0.09 }$&0.21$\pm^{ 0.04 }_{ 0.06 }$&$<$0.279\\ 
2729&5267$\pm^{ 197 }_{ 54 }$&4.55$\pm^{ 0.10 }_{ 0.03 }$&-0.33$\pm^{ 0.32 }_{ 0.33 }$&-0.14$\pm^{ 0.09 }_{ 0.02 }$&-0.13$\pm^{ 0.03 }_{ 0.03 }$&$<$0.277\\ 
2735&5127$\pm^{ 328 }_{ 112 }$&2.52$\pm^{ 0.36 }_{ 0.36 }$&-0.66$\pm^{ 0.72 }_{ 0.50 }$&0.20$\pm^{ 0.25 }_{ 0.24 }$&1.06$\pm^{ 0.19 }_{ 0.13 }$&$<$0.273\\ 
2898&4813$\pm^{ 89 }_{ 51 }$&4.64$\pm^{ 0.02 }_{ 0.08 }$&0.18$\pm^{ 0.21 }_{ 0.35 }$&-0.09$\pm^{ 0.03 }_{ 0.06 }$&-0.15$\pm^{ 0.03 }_{ 0.03 }$&$<$0.273\\ 
2944&6168$\pm^{ >831 }_{ 132 }$&4.03$\pm^{ 0.29 }_{ 0.09 }$&-0.30$\pm^{ 0.51 }_{ 0.30 }$&0.02$\pm^{ 0.17 }_{ 0.05 }$&0.22$\pm^{ 0.04 }_{ 0.08 }$&$<$0.274\\ 
2957&$<$4600&\ldots&\ldots&\ldots&\ldots&\ldots\\ 
2961&5909$\pm^{ 573 }_{ 225 }$&4.12$\pm^{ 0.25 }_{ 0.13 }$&-0.12$\pm^{ 0.60 }_{ 0.42 }$&-0.00$\pm^{ 0.17 }_{ 0.07 }$&0.16$\pm^{ 0.04 }_{ 0.07 }$&$<$0.282\\ 
2995&5452$\pm^{ 149 }_{ 112 }$&4.55$\pm^{ 0.03 }_{ 0.13 }$&0.27$\pm^{ 0.21 }_{ 0.39 }$&0.00$\pm^{ 0.03 }_{ 0.09 }$&-0.05$\pm^{ 0.03 }_{ 0.03 }$&$<$0.280\\ 
\hline
\end{tabular}
\end{center}
\label{short_cat}
\end{table*}%
\clearpage
 \appendix
 \section{Matching with Gaia data}
 \label{gaia_section}
  \begin{table*}[h]
\caption{A short example table of our matched Gaia detections for TESS CTL objects. All astrometric data is from Gaia. ``Match type" indicates which of the category that describes the match of each object to its Gaia counterpart, 1. only magnitude match within 1" of the proper motion corrected position, 2. best magnitude match within 1" of the proper motion corrected position, 3. only magnitude match within 3" of the proper motion corrected position, 4. best magnitude match within 3" of the proper motion corrected position, 5. only magnitude match within 5" of the proper motion corrected position, 6. best magnitude match within 5" of the proper motion corrected position, 7. only magnitude match with a two parameter astrometric solution within 5" of the uncorrected position, 8. best magnitude match with a two parameter astrometric solution within 5" of the uncorrected position.}
\tiny
\begin{center}
\begin{tabular}{rrrrrrrrrrc}
\hline
TESS ID&GAIA DR2 ID&R.A.&Dec.&$\mu_{\alpha}\cos\delta$&$\mu_{\delta}$&$\pi$&G&R.V.&RUWE&match\\
&&\multicolumn{2}{c}{(Ep.=2015.5 Eq.=J2000)}&(mas/yr)&(mas/yr)&(mas)&(mag.)&(km/s)&&type\\
\hline
101&6220284483485052800&218.753426&$-$29.786162&  $-$7.70$\pm$   0.08&  $-$6.00$\pm$   0.06&  3.38$\pm$  0.04&11.676&$-$24.190$\pm$ 3.130&0.850&1\\ 
159&6220293971069949696&218.768146&$-$29.677842&   0.49$\pm$   0.12&  $-$2.69$\pm$   0.17&  2.56$\pm$  0.07&12.207&$-$17.790$\pm$ 0.920&0.958&1\\ 
164&6220294039789428864&218.768010&$-$29.665272&  $-$3.93$\pm$   0.07& $-$14.69$\pm$   0.09&  2.26$\pm$  0.04&13.362&\ldots&1.034&1\\ 
185&6220295448538709376&218.742945&$-$29.629234& $-$18.13$\pm$   0.11& $-$31.40$\pm$   0.13&  3.32$\pm$  0.06&12.114&$-$1.250$\pm$ 1.400&1.029&1\\ 
358&6221815798241951872&218.798872&$-$29.348655&  $-$9.77$\pm$   0.11&  $-$9.79$\pm$   0.14&  3.12$\pm$  0.07& 9.584&$-$11.480$\pm$ 1.940&0.860&1\\ 
382&6221815970040649728&218.831883&$-$29.314487&   2.23$\pm$   0.08& $-$12.55$\pm$   0.08&  3.48$\pm$  0.05&10.565&19.120$\pm$ 0.510&0.989&1\\ 
479&6221819577813206656&218.834572&$-$29.156620&  $-$3.53$\pm$   0.05&  $-$5.22$\pm$   0.04&  1.12$\pm$  0.02&13.381&21.100$\pm$ 5.970&1.020&1\\ 
733&6222034291818114688&218.731298&$-$28.790260&  38.48$\pm$   0.16&  18.49$\pm$   0.15&  4.29$\pm$  0.10&16.322&\ldots&0.904&1\\ 
880&6222112460222994560&218.727346&$-$28.503355& $-$74.65$\pm$   0.20& $-$12.53$\pm$   0.15&  2.19$\pm$  0.10&17.116&\ldots&0.897&1\\ 
1068&6222126405981353472&218.798071&$-$28.235262& $-$65.49$\pm$   0.50& $-$34.51$\pm$   0.40&  4.79$\pm$  0.34&18.096&\ldots&1.038&1\\ 
\hline
\end{tabular}
\end{center}
\normalsize
\label{short_gaia}
\end{table*}%
 We used the VO Multicone search available in TopCAT \citep{Taylor2005} to select possible Gaia matches for objects in the TESS CTL. The TESS Input Catalog uses positions corrected to Epoch 2000.0. As we are comparing to Gaia which has Epoch 2015.5, we need to take into account the possible motions of stars since then when searching for matches. Hence we set our search radius for each object as twenty times the total proper motion in arcseconds per year listed in the TESS CTL or ten arcseconds, whichever was greater. This both allows us to search for fast-moving stars and to have a buffer to allow for errors in the position and proper motion. 
 
To ensure that we were matching to the same star and not a coincident, likely fainter object, we needed to know the Gaia magnitude of each TESS source. The TESS CTL lists Gaia DR1 magnitudes for most objects as well as estimated TESS $T$ magnitudes for all objects. We used these as our Gaia $G$-band magnitude when available. The Gaia DR1 and DR2 $G$-band magnitudes have a small colour term between them of around 0.25\,mag for the coolest stars \citep{Brown2018}. To take this into account we set a magnitude difference tolerance of one magnitude between the Gaia DR2 $G$-band magnitude and the Gaia $G$-band magnitude listed in the CTL. For redder objects ($T-J>1.1$) and brighter objects ($T<7$) we increase this tolerance to two magnitudes. Where no $G$-band magnitude is listed in the CTL we use the following relation, calculated from the Gaia DR1 $G$, TESS $T$ and 2MASS $J$ magnitudes of 935 randomly selected TESS targets,
\begin{equation}
 T-G=0.14666639-1.3241485(T-J)+0.32423058(T-J)^2
\end{equation}   
All objects in the TESS CTL have a $T$ magnitude and almost all have a $J$ magnitude. Hence this allows us to calculate Gaia DR1 $G$-band magnitudes for sources with no $G$-band magnitude listed. For objects where $T-J$ fell outside the typical range (between 0.0 and 2.0) we also increased our magnitude tolerance to two magnitudes.

We then follow a multi-step approach to matching. Firstly we must calculate corrected positions for all potential Gaia matches to the CTL epoch of 2000.0. Then we followed the process below
\begin{enumerate}
\item If there were matches within 1" of the corrected position and within the $G$-band magnitude tolerance then select the match with the smallest magnitude difference to the CTL magnitude.
\item If not then repeat the process above but with a match radius of 3".
\item If not then repeat the process above but with a match radius of 5".
\item If there are no matches within 5" of the corrected position then find any object which only has a Gaia two parameter solution and a matching magnitude. If there are multiple two parameter solution matches then select the one with the smallest magnitude difference to the CTL magnitude.
\end{enumerate}
The vast majority (3,731,686/3,817,768) of our sources only have one position and magnitude match within one arcsecond. The two parameter object search is to take into account objects which appear in the TESS CTL but which do not have Gaia DR2 proper motions or parallaxes. While this lack of a parallax precludes us from including these 60,856 objects in our stellar parameters analysis, we include them here for completeness. We include the Reduced Unit-Weight Error (RUWE) \citep{Lindegren2018,Lindegren2018a} from the ARI GAVO DR2light table. This can be used to filter out noisy astrometric solutions.

Table~\ref{short_gaia} is a short example of our full electronically available TESS-GAIA match.

\end{document}